\documentclass[nofootinbib,twocolumn,5p]{revtex4} 

\usepackage{amsmath}
\usepackage{graphicx}
\usepackage{graphics}
\usepackage{dcolumn}
\usepackage{bm}
\usepackage{color}
\usepackage[normalem]{ulem}

\usepackage[pdftex]{hyperref}
\usepackage{graphicx} 
\usepackage{epstopdf} 
\usepackage{dcolumn}
\usepackage{amssymb}
\usepackage{amsmath}
\usepackage{fixltx2e} 
\usepackage{pict2e} 

\begin{document}

\title{The effect of social balance on social fragmentation}

\author{Tuan Minh Pham$^{1,2}$, Imre Kondor$^{2,3}$, Rudolf Hanel$^{1,2}$, Stefan Thurner$^{1,2,4,5}$}
\email{stefan.thurner@meduniwien.ac.at} 

\affiliation{
$^1$ Section for the Science of Complex Systems, CeMSIIS, Medical University of Vienna,  Spitalgasse 23, A-1090, Vienna, Austria\\
$^2$ Complexity Science Hub, Vienna Josefst{\"a}dterstrasse 39, A-1090 Vienna, Austria \\
$^3$ London Mathematical Laboratory, 8 Margravine Gardens, Hammersmith, London W6 8RH, UK\\
$^4$ Santa Fe Institute, 1399 Hyde Park Road, Santa Fe, NM 87501, USA\\
$^5$ IIASA, Schlossplatz 1, 2361 Laxenburg, Austria 
} 

\begin{abstract}
With the availability of cell phones, internet, social media etc. the interconnectedness of 
people within most societies has increased drastically over the past three decades.
Across the same timespan, we are observing the phenomenon of increasing levels of fragmentation in society into 
relatively small and isolated groups that  have been termed filter bubbles, or echo chambers. 
These pose a number of threats to open societies, in particular, a radicalisation in political, social or cultural issues, 
and a limited access to facts. In this paper we show that these two phenomena might be tightly related. 
We study a simple stochastic co-evolutionary model of a society of interacting people.   
People are not only able to update their opinions within their social context, but  can also update
their social links from collaborative to hostile, and vice versa. 
The latter is implemented such that social balance is realised. 
We find that there exists a critical level of interconnectedness, above which society fragments into small sub-communities that are 
positively linked within and hostile towards other groups. 
We argue that the existence of a critical communication density is a universal phenomenon  
in all societies  that exhibit social balance. 
The necessity arises from the underlying mathematical structure of a phase transition phenomenon that is known from 
the theory of a kind of disordered magnets called spin glasses. 
We discuss the consequences of this phase transition for social fragmentation in society. 

\end{abstract}

\maketitle

\section{Introduction}
\label{sec:introduction}

Social cohesion and social fragmentation are central topics  in
the organisation and functioning of large-scale societies. 
As such it is a central topic in sociology since its very beginning. 
Starting with Durkheim \cite{Durkheim}, who referred to the mutual dependencies between individuals 
as ``organic solidarity'', the concept of social cohesion has evolved, however, it remains a core theme in sociology 
\cite{pahl1991,Bruhn2009, Dubet2013}. 
Over the past two decades, 
concerns have been raised that modern societies might  gradually be losing their cohesion 
\cite{Stanley,  Chan, council2008report, Jenson, OECD2011, dragolov2016}.  
This has been attributed to several ongoing changes: globalisation, migration and ethno-cultural diversity, 
modern communication technologies, and the integration of states into trans-national entities, 
such as the European Union \cite{Schiefer}. 
As the cohesion of a society declines, it faces the threat of becoming fragmented, 
which might come with a number of potentially catastrophic consequences, 
such as riots, civil wars,  governmental shutdowns, or the  decline of democracy \cite{denton2016}. 
Hence it has become a great challenge of how to preserve social cohesion without interfering with diversity 
\cite{Berger, Gough}.

Despite the lack of a consensus on what constitutes social cohesion, 
social relations  have been widely regarded  among the 
most essential aspects \cite{Schiefer}. 
Both, social  cohesion and fragmentation, 
 emerge from complex interactions between individuals. 
One mode of collective social organisation can change to another if interactions change: 
individuals initially united by cooperation for a common good can become segregated  
once they start competing for their  ethno-cultural,  economical, or political values (or identities) \cite{Fukuyama}.
In many societies transitions between  fragmented and cohesive ``phases''  happen throughout history \cite{Bodnar}. 
In line with this view, here we define fragmentation as the regime (phase) in which society-wide  
collaborative efforts are broken down into local cooperation within subcommunities, 
with little or no collaboration between these groups. 

\subsection{New social media and social fragmentation}

Local interactions between individuals shape and define the nature and quality of the overall social organisation. 
Novel communication technologies affect both the quality and the quantity of social interactions 
and thus might have a crucial impact on social cohesion.  
Among  these new possibilities the effect of 
social media on  social cohesion has been studied \cite{Bright}. 
On the one hand,  social media may create so-called {\em echo chambers} in political discourse \cite{Kiran,  delvalle}, 
where individuals reinforce their current position by repeated interactions with those of the same view.  
On the other hand, social media increasingly guide individuals to contents they are likely to agree with,  resulting in the danger of so-called {\em filter bubbles} \cite{pariser2011, Flaxman}.  
These phenomena might play an important role in the radicalisation of political discourse  
and the  decline of cross-ideological  exposure -- one of the building blocks of democracy \cite{Sunstein}. 
Recently, several models have been proposed  to better understand the formation of echo chambers 
and filter bubbles, as well as their effects on fragmentation \cite{Evans, Sirbu2019, Baumann}.

\subsection{Modelling of social fragmentation} 

Modelling social fragmentation has a long history \cite{Bramson}. In the context of cultural dynamics,  
Axelrod predicted fragmentation into cultural groups if individuals within a given ``social neighbourhood''  
are more likely to interact with similar others than with dissimilar ones \cite{RAxelrod1997}.  
The more they interact with each other, the more similar  they  become and thus 
the higher the chance  is for future interactions. 
The dynamics continues until stable regions of identical individuals are established.  
However, the formation of such cultural regions is observed only for small societies; large societies reach global consensus. 
Recent research pointed out two properties of the Axelrod model, namely,  
the fragility of the fragmented phase with respect to random perturbations, 
such as the ``mutation'' of cultural features \cite{Klemm2003a}, 
and the fact that the transition from complete homogeneity to cultural diversity 
only happens beyond a critical number  of alternative traits per feature \cite{Castellano2000}. 
These shortcomings were later resolved by replacing the interpersonal influence in the original 
model by social influence   \cite{Flache2011, Battiston2017}. 
In the context of segregation, Schelling's celebrated model for the distribution of people of different races, 
assumes that individuals prefer to be in a neighbourhood with the majority of their own type \cite{Schelling}.  
Complete segregation into clusters of one type 
 occurs as they move from one neighbourhood to 
another to satisfy their preferences. Subsequent research has shown that this way to understand segregation is 
quite robust \cite{pancs}.  Axelrod's and Schelling's models both explain  fragmentation (segregation) 
as an emergent  collective phenomenon that results from  individuals' incentives only. 
A question that remains open, however, is how the transition from cohesion to fragmentation   
corresponds to the rearrangement of social ties.

\subsubsection{Theory of social cohesion -- structural balance} 

\begin{figure}[tb]
\begin{center}
\includegraphics[scale=0.28]{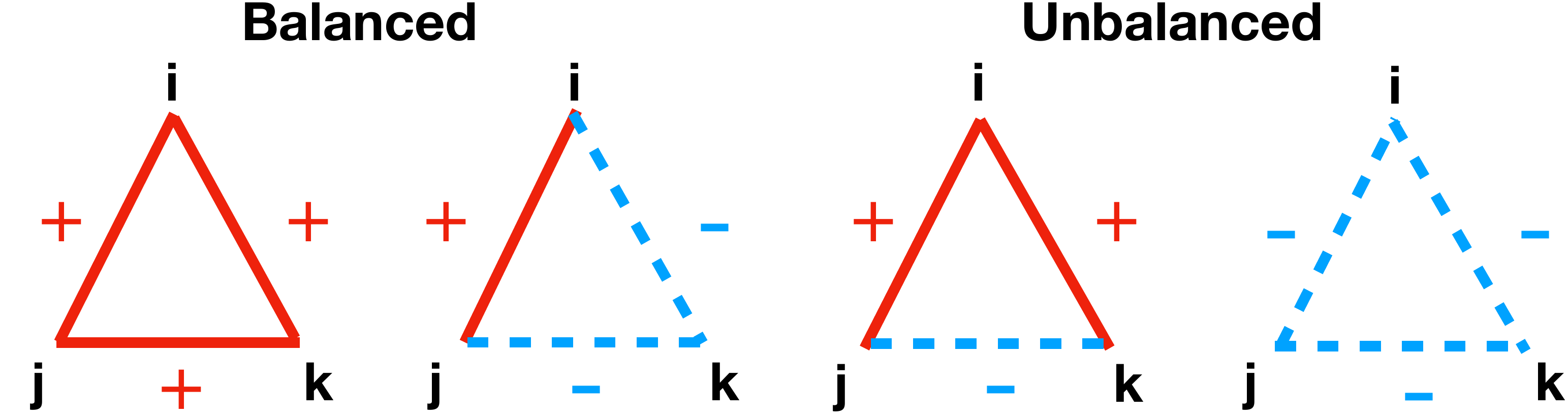}
\end{center} 
\caption{Balanced and unbalanced triangles.  
Red lines denoted with a plus sign represent friendly and cooperative relations between individuals $i$, $j$, and $k$.
Dashed blue lines (minus) are negative or hostile links. A link between node $i$ and $j$ is denoted by $J_{ij}$. It can be 
$J_{ij}=+1$ or $J_{ij}=-1$.
A triangle is called balanced, if the product of its three link states is $J_{ij} J_{jk} J_{ki} = 1$,
and unbalanced, if the product is $J_{ij} J_{jk} J_{ki} =-1$. The first two triangles are balanced, while the second two are not.}
    \label{fig:triangles}
\end{figure}

One of the seminal ideas in sociology of the 20th century was the concept of \textit{structural balance}, 
which is based on  the observation that social dynamics in cliques of individuals is determined less by pair-wise, 
but by triadic relations, i.e. triangle relations become the more fundamental unit. 
Structural balance theory was first proposed by Heider in the 
1940s \cite{Heider} and states that a group of three individuals forms a {\em balanced} triangle, if either all 
the three are mutually friends (positive relation) or two of them are friends and both have the same enemy (negative link). 
Three people form an {\em unbalanced} triad or triangle, if either all the three are mutual enemies, 
or if two of them are enemies but the third is their mutual friend, see Fig. \ref{fig:triangles}. 
Empirically, balanced triangles are found much more frequently than unbalanced ones in human societies, 
for a list of recent results, see e.g. 
\cite{Hummon, Szell, Szell2010, Szell2012, Leskovec, Facchetti2011, Facchetti2012, estrada, Zheng, Sadilek2018, Kirkley, Lerner, Neal}. 
If an unbalanced situation occurs, individuals seem to strive to eliminate the associated tension 
by flipping the sign of one of the three links, resulting in a balanced arrangement. 
A perfectly balanced society would be one in which there are no unbalanced triangles --  all individuals enjoy life 
without any tension. 
In contrast, a cohesive but not stress-free society is more conducive to change and/or improvement.

\subsubsection{Our definition of social fragmentation}

From what has been discussed so far it is clear that one must distinguish 
between different concepts of social fragmentation: 
urban fragmentation, segregation, social balance, loss of coherence because of evaporation of joint ideals, etc. 
These concepts all capture various aspects of social cohesion. 
For the following, we are interested in a broad and generic definition of social fragmentation, 
closely following Heider's notion of social balance: 
We call a society {\em fragmented} if there are many groups that are locally collaborative with 
a high density of ``positive links'' within the group, but are often hostile to other groups. 
On the other hand, a society is  {\em cohesive}  if one finds a sufficient density of positive links 
between groups, such that one can ``travel''  from group to group, without ever 
having to use negative links. In other words, we define a society as cohesive, 
if the positive links percolate. 

\subsubsection{Co-evolution and adaptive networks} 

Collective dynamics of social systems has been studied  within the framework of {\em adaptive networks}; 
for an overview see e.g. \cite{thurnerbook}. 
In this approach, fragmentation results from a co-evolutionary rearrangement of social ties, 
together with updates of individual traits (``states'') \cite{Bohme}.                                                                                                                                                                                                                                                                                                                               In \cite{Holme}, new relationships are created between people of the same opinion by rewiring pre-existing connections 
with a given probability. As  the rewiring rate increases, the system self-organises into many communities, 
such that members of the same community converge on their opinions, but strongly differ from members of other groups. 
In a modification of the original Axelrod model \cite{Centola}  links between dissimilar agents (with no feature 
in common) are replaced by links  between agents who may be either similar or dissimilar. 
This mechanism was shown to  change the network structure from a regular lattice  to a network with multiple clusters 
and  to significantly increase the critical point of the transition from a mono- to a multicultural regime. 
In \cite{Kozma} only links between agents whose opinions differ from each other more 
than a tolerance level may be broken. New links are established to other agents, regardless of their opinions.   
As a consequence, the transition from consensus towards fragmentation happens at a lower tolerance level 
than on fixed networks.

\subsubsection{Earlier models on opinion formation}
 
Many previous approaches towards  modeling  social dynamics  focused on a setting where individuals are 
characterised by a number of socio-economic traits. These become  
dynamical variables  and one cam study their collective evolution with numerical methods. 
This direction has shaped the field of {\em opinion dynamics}  \cite{Castellano, Sirbu2017}. 
Influential models in this field are DeGroot's model of belief consensus \cite{DeGroot}, 
the voter model \cite{votermodel, Holley}, and the majority-rule model \cite{Galam2002}. 
A generic and unrealistic feature of these models is that generally global consensus is  
established among the agents, regardless of the  detailed dynamics or the underlying network structures. 
There are, however, models that do exhibit either consensus or opinion fragmentation. 
They either rely on the  ``bounded confidence'' assumption that states that only those 
whose opinions differ less than a given level can interact \cite{Deffuant, Hegselmann2002}, 
or they employ  the fact that individuals only adopt their views once 
a certain fraction of their neighbours did \cite{Watts2002, Klimek2008}. 
Both types of models show a phase of global consensus if the confidence level (fraction of neighbours) 
exceeds a critical value. Below this threshold, clusters of different opinions  of various sizes  appear. 

\subsubsection{Opinion dynamics on signed networks}

The first attempt to incorporate Heider's social balance into opinion dynamics was in \cite{Altafini, Altafini2013, Altafini2015}. 
They showed that an opinion formation process on a balanced network ends up in polarised states, 
where contradictory opinions are clustered into two  groups. 
This result was extended to the case of  time-varying signed graphs in  \cite{Proskurnikov, Meng},  however, 
there  opinion- and  network dynamics are not coupled.  
In yet another class of models that is based on the Hebbian learning rule \cite{Hebb} the weight of the social  
link between two individuals is assumed to be a function of the correspondence between their states. 
As their opinions evolve over time, the weight increases (decreases) proportional 
to their opinion concordance (discordance). 
Fragmentation has been
 shown to emerge from such adaptive dynamics, \cite{Macy,Singh}. 
There the network only reacts to the change of opinions, but does not emerge from Heider's principle 
of minimising social tension. Saeedian et al. \cite{Saeedian2019} recently consider a co-evolutionary 
dynamics where not only  friends with opposite opinions but also  enemies with similar ones can change either their opinions 
or their relations to remove cognitive dissonance. In the final frozen states, 
the network fragments into groups of friendly and like-minded individuals who, however, are hostile to members of the other groups.
\\

In this paper we propose to understand the mechanism of social fragmentation as a consequence of social balance. 
To this end we study a minimalistic stochastic, co-evolutionary model where individuals tend to avoid social stress 
by either adopting their opinions, or by changing their social links from cooperative to hostile, or vice versa. 
Heider's concept of social balance is explicitly taken into account by  co-evolutionary evolution mechanism,  
rather than  being  imposed {\em a priori} or emerging from the  dynamics of the social network alone.
We will see that the model allows us to understand the emergence  of echo chambers and 
filter bubbles as a function of the average connectivity of the society. 
We find a fundamental regime shift (or phase transition) that happens at critical values of social connectivity.
Below the critical density we observe a largely cohesive society,  
above it there exists an unavoidable phase that is dominated by the existence of many small  
collaborative communities, characterised by hostile links towards other groups. 

\section{The model}
\label{sec:model}

\subsection{A co-evolutionary model of opinion- and social network formation}

\begin{figure}[tb]
\begin{center}
\includegraphics[scale=0.29]{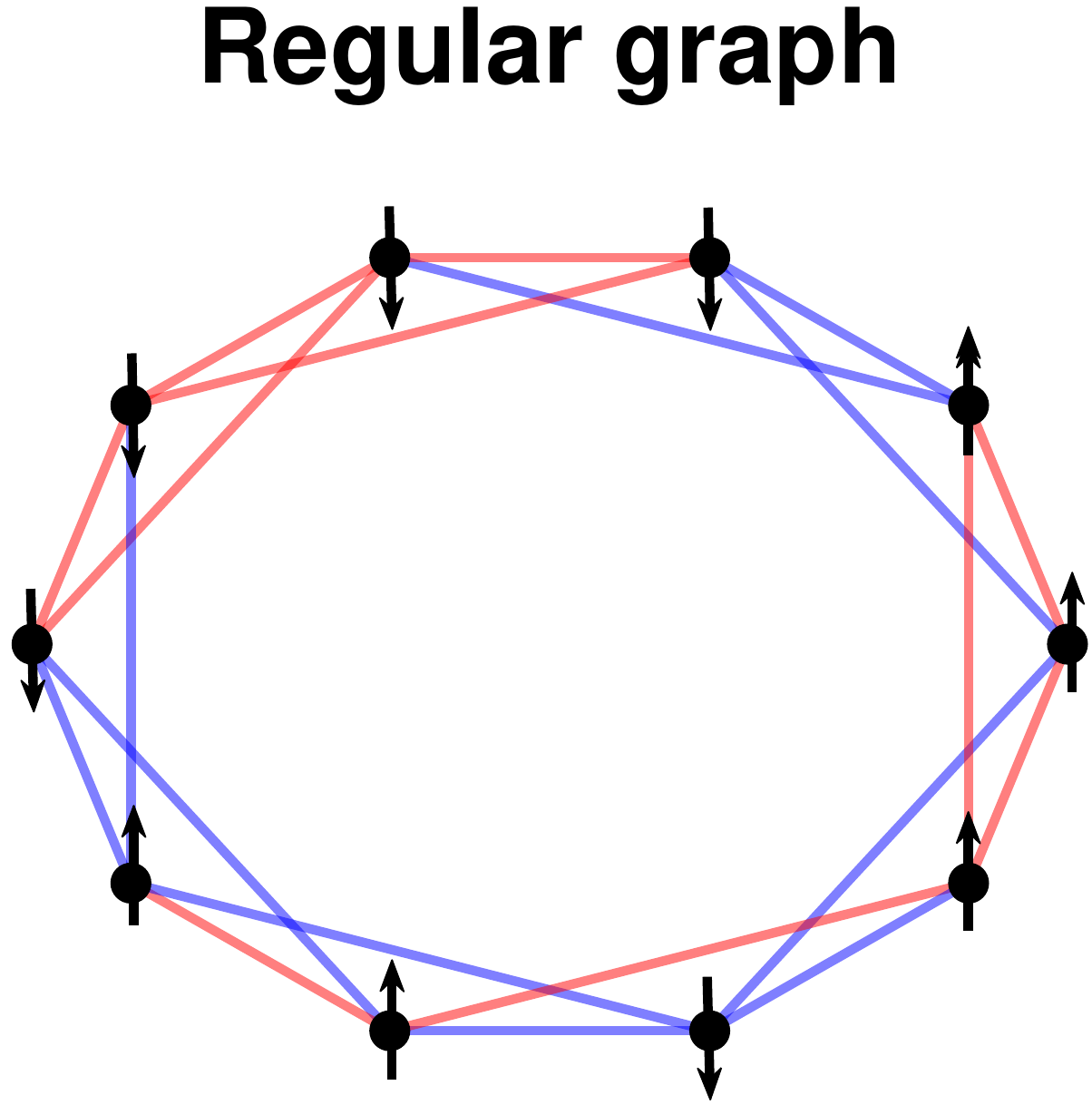}
\qquad
\includegraphics[scale=0.29]{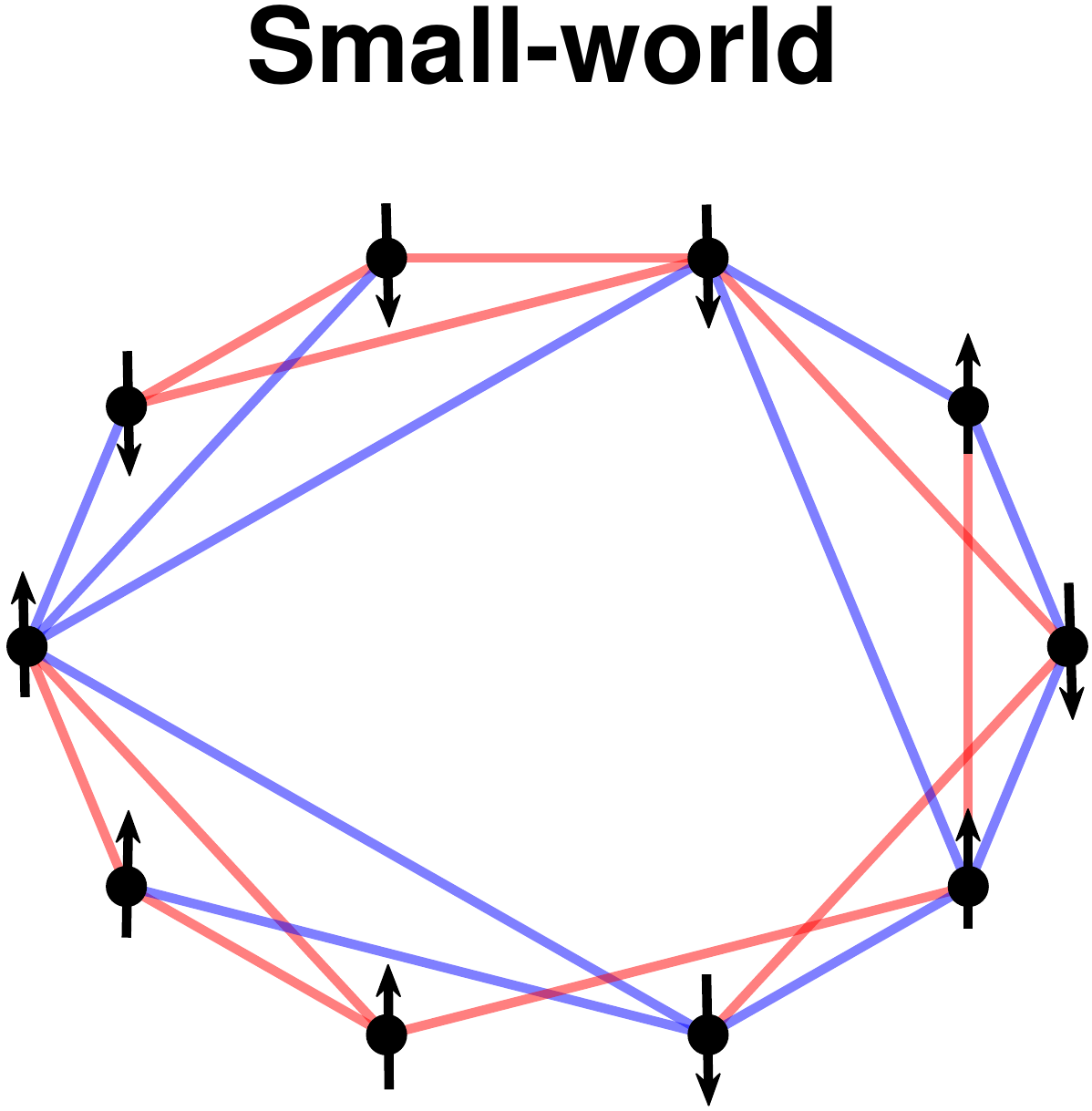}
\end{center} 
\caption{Network structure of our model society. 
Nodes represent individuals who have binary opinions that are displayed as either $\uparrow$ or $\downarrow$. 
Individuals are either linked by positive  (red) or negative (blue) social ties. 
(a) shows a regular network topology, i.e. every node has the same number of neighbours; here $k=4$. 
In (b) nodes are linked to others in a small world network that can be obtained from (a) by randomly rewiring 
one side of any link  with probability of $\epsilon = 0.2$. 
Not everyone has the same number of neighbours anymore. 
} 
   \label{fig:networks}
\end{figure}

We assume that a society is made up by $N$ individuals that we label by latin indices, $i$. 
Each of these individuals is embedded in a social network and has social relations to $k_i$
fellow individuals that are labelled by $j$. 
We keep the average number of links per person $k = \bar  k_i$ as a model parameter. 
This number is assumed to be fixed over time. Each relation
between $i$ and $j$ can be either positive, $J_{ij}=1$, e.g. if they are friends, or negative, 
$J_{ij}=-1$, if they are enemies.  
If two  individuals are linked with a negative link this indicates a certain level of social stress. 
Each individual is endowed with an opinion, $s_i$. For simplicity we assume that there exists only one 
type of binary opinion, of the type: yes or no,  Trump or Hillary, etc. 
In Fig. \ref{fig:networks} we show a schematic picture of our model society in the simplest case,  
where a total of $N$ individuals with opinions ($\uparrow$ and $\downarrow$) are linked to  $k$ neighbours each 
in a regular way (a)  and in a so-called {\em small world} network  \cite{WattsStrogatz} (b) with the same average connectivity, $k$.  

Imagine that two individuals are linked through a positive link and they have opposite opinions on a given 
subject. We assume that this will cause a certain amount of social stress in the system. 
If, on the contrary, the two individuals do not like each other, $J_{ij}=-1$, and they have opposite opinions, this will not 
lead to additional social stress.  
Both, the opinions and the quality of the social links, can be updated. 
Whenever  $i$ changes her opinion, we have $s_i= 1 \to s_i= -1$, or $s_i= -1 \to s_i= 1$. 
The same is true for social links, whenever we change friendship to enmity, $J_{ij}= 1 \to J_{ij}= -1$, or vice versa.
We assume that on average individuals tend to update their opinions and social links   
such that they reduce their local levels of social stress. 
To keep track of the total amount of social stress, we introduce a function, $H$, which allows 
us to formulate a simple stochastic co-evolutionary model.

\subsection{Minimising social stress -- a Hamiltonian approach}

The system under study evolves to minimise overall social tension, which can be defined as
\begin{equation}
H =- \sum_{(i,  j)} J_{ij} s_i s_j \,\,- g\, \sum_{(i, j, k)} J_{ij} J_{jk} J _{ki} \;,
\label{eq:model}
\end{equation}
where $s_i \in \{-1,1\}$ denotes the opinion of an individual $i$ and  $J_{ij} \in \{1,-1\}$ represents friendship and enmity  between  two connected agents $i$ and $j$, respectively ($J_{ij} = 0$, if they are not linked). This type of cost function is called a {\em Hamiltonian} function in physics, where it captures the total energy in a system as a function of its configuration, $H = H(s_i, J_{ij})$. 
There it is then used to implement the principle of minimisation of energy. 

\begin{figure*}[ht]
\begin{center}
\includegraphics[scale=0.19]{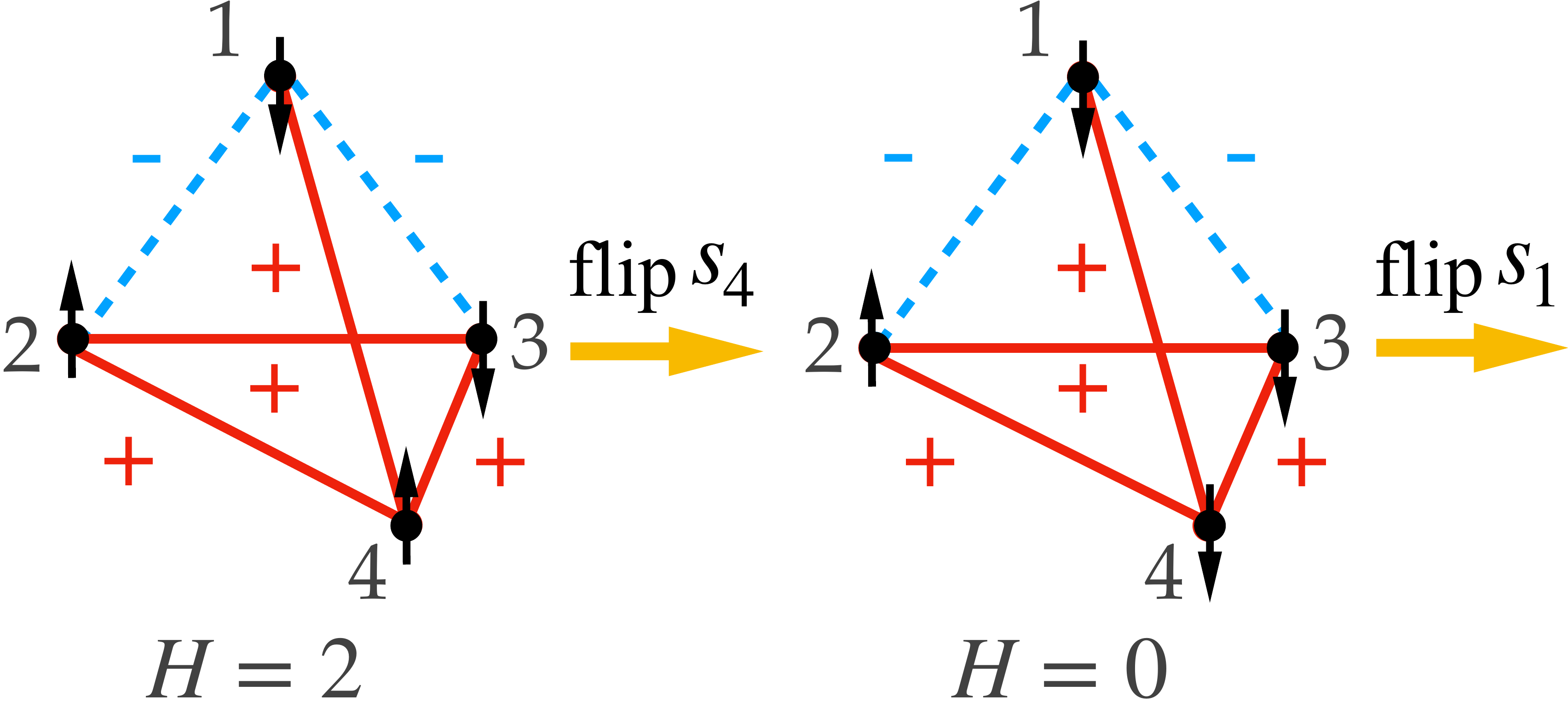}
\includegraphics[scale=0.19]{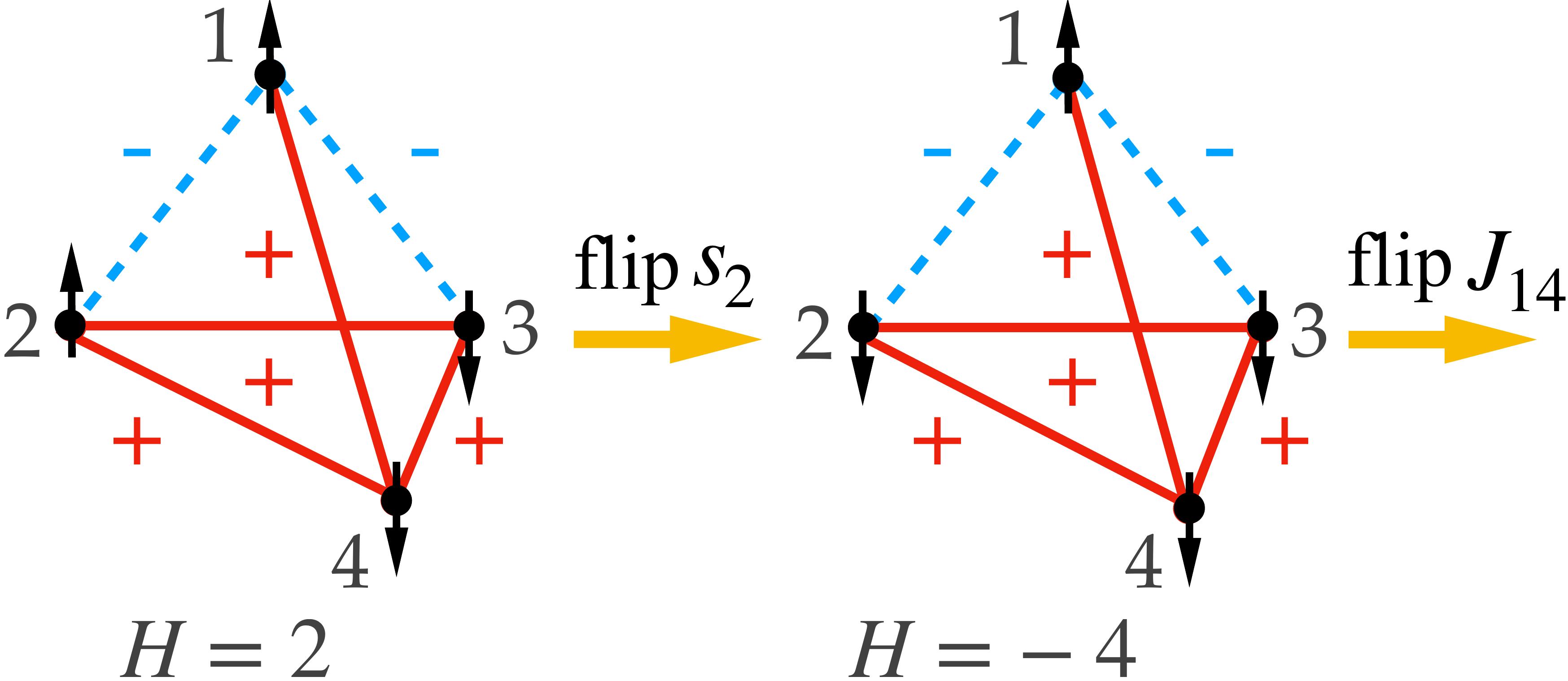}
\includegraphics[scale=0.19]{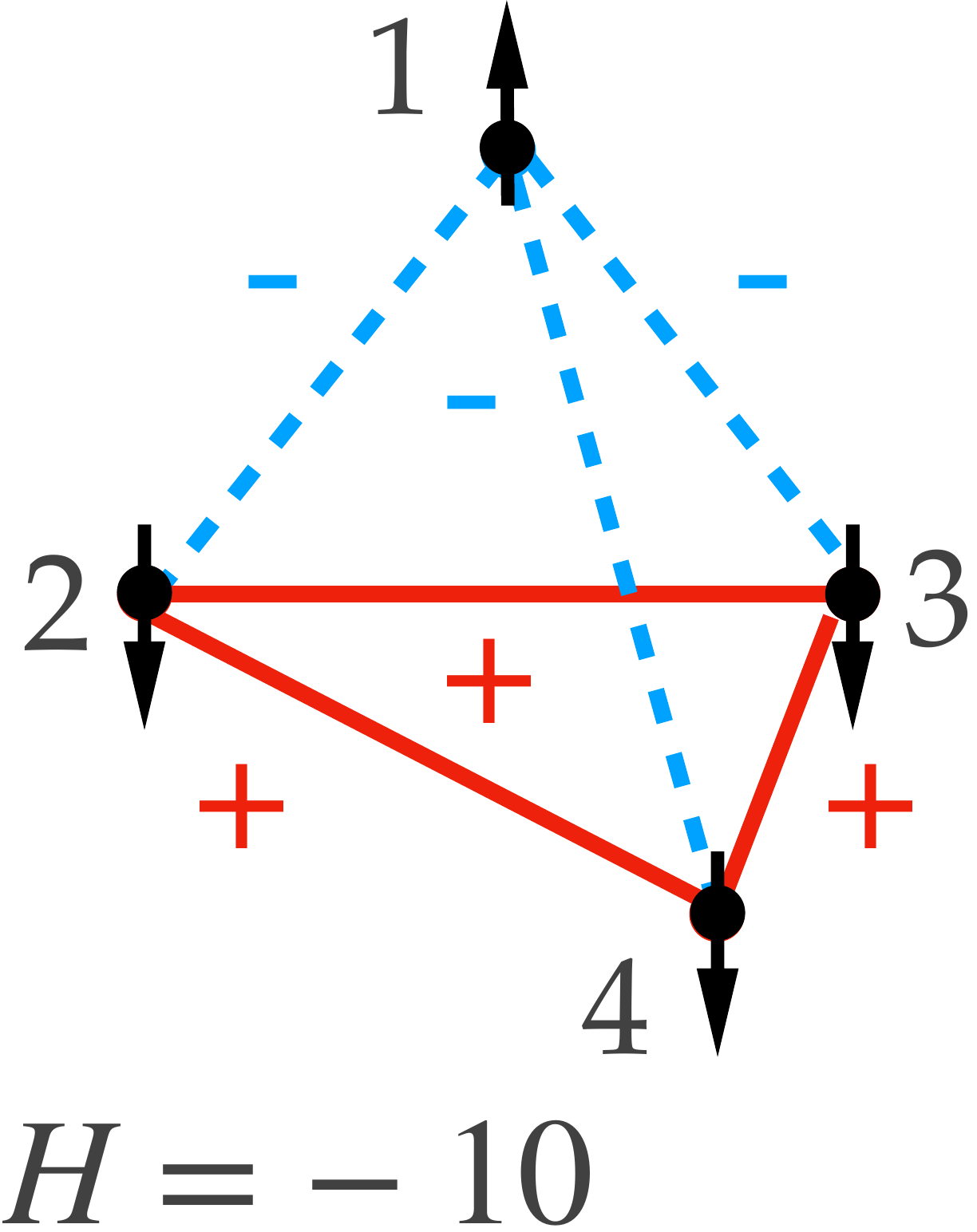}
\vskip0.3cm
\end{center}
\caption{Example for how social stress, $H$, is gradually lowered in a sequence of changes of opinions and  links 
between four individuals (assuming $g=1$ for simplicity). 
Opinions of the nodes are given by $\uparrow$ and $\downarrow$, positive links are red, negative are blue. 
In the course of this sequence the number of balanced triangles changes from two to four. 
Note that in the second step social stress is temporarily increased. 
This is a consequence of the stochastic nature of the model, where also unfavourable events happen from time to time.  
}
    \label{fig:evolution}
\end{figure*}

In Eq. (\ref{eq:model}), the first sum describes the opinion adoption process between interacting agents. 
It assumes that individuals should act in such a way as to avoid cognitive dissonance among them:  
if $i$ and $j$ are friends, they are more likely to share the same view,  otherwise, they may hold opposing opinions. 
Following the ``social influence" theory  by \cite{Festinger, French, Harary1959}, for any individual $i$,  
the simultaneous influences from all its neighbours are represented by the sum over $j$ of $(J_{ij}s_j)s_i$ terms.
The second term explicitly takes care of Heider's social balance: 
it incorporates the tendency of suppressing unbalanced triangles between individuals. 
This effect is implemented by the sum over {\em all} possible triadic relations between any three individuals 
$i$, $j$, and $k$. If $J_{ij}J_{jk}J_{ki}  = 1$, they feel no social tension, otherwise social balance pushes them to switch their relations. 
Note that a link between $i$ and $j$, $J_{ij}$, in general will belong to several triangles. A flip of $J_{ij}$
that lowers the {\em total} number of unbalanced triads should happen with a higher probability than a flip 
that leads to an increase of unbalanced triangles, i.e. increases overall social stress. 
See the next subsection for how this is implemented.
The parameter $g$ in Eq. (\ref{eq:model}) controls the  relative strength of the social balance term 
with respect to the opinion formation contribution (first term). 
In accordance with Heider's theory, $g$ must be positive so that balanced triangles 
do indeed  dominate the unbalanced ones\footnote{
At this point  it is not clear how to empirically infer the value of $g$ for a given society. 
We chose $g \in [0, 1]$ in the model implementation. In fact, any non-negligible value, $g \neq 1$, 
can be shown to yield similar results as $g = 1$ (see SI).   
This choice corresponds to the assumption that the effect of social balance is comparable in importance to the 
opinion terms (social influence) in Eq. (\ref{eq:model}).  
As long as the number of links and that of triangles are of the same order of magnitude (as is the case in sparse networks),  
it seems reasonable to keep the contribution of  the Heider term comparable to that of opinions.
In physics the case $g = 0$ corresponds to the classical  Edwards-Anderson spin glass model  \cite{Edwards1975}, 
while the other extreme, $g \rightarrow \infty$,  corresponds to the model studied in  \cite{Antal2005, Marvel,  Rabbani}.  
}.

Figure \ref{fig:evolution} shows an example for how four individuals with  given initial opinions and 
links can change social stress, $H$, by flipping either opinions or links. 
The configuration starts with a situation that amounts to $H=2$. When node 4 flips its opinion from $\uparrow$ to $\downarrow$, 
social stress is decreased to $H=0$. 
Next, node 1 flips its opinion to $\uparrow$ and increases  stress to $H=2$. This is not what happens usually, but since the 
dynamics is stochastic, these situations will also occur. 
In the next step,  node 2 flips from $\uparrow$ to $\downarrow$ and thereby lowers social stress to  $H=-4$.
Finally, by 1 and 4  flipping their link from positive to negative, we arrive at a relatively stress-free situation, $H=-10$, 
that is socially balanced.

\subsection{A stochastic co-evolutionary model -- the Metropolis algorithm}

The social stress function, $H$, now specifies the way by which the dynamical variables, 
$s_i$ and $J_{ij}$, change over time.  
Assuming that humans generally tend to reduce social stress, changes that decrease  $H$ are favoured over those increasing it. 
We implement the joint evolution of opinions and links by the so-called Metropolis algorithm \cite{Metropolis}.  
Starting from a random configuration of opinions and links, the society is updated from one timestep $t$ to the next as follows:

\begin{enumerate}
\item Compute $H$ of the current system, assume it has a value of $H_0$.

\item Pick a node $i$ at random and flip its opinion, $s_i$. Compute $H$ again, it is now $H_1$. 
If the value of $H$ has decreased in response to the flip, $H_1 \leq H_0$, accept the flip. 
If the value of $H$ increased, accept the flip only with probability, $p = e^{  - \Delta H/T } $, where  
$\Delta H=H_1-H_0$ is the difference of stress before and after the flip. 
$T$ denotes the ``social temperature" and is a model parameter.
Pick the next node randomly and continue until $N \times n$ opinion updates have been performed (Monte Carlo iterations).

\item Compute $H$ of the system at this point, assume that it is now $\tilde{H}_0$. 
We now pick one link randomly,  $J_{ij}$,  and flip it. Compute $H$ again, and assuming it to be $\tilde{H}_1$, we 
accept the flip if $\tilde{H}_1 \leq \tilde{H}_0$, and accept it with probability $p' = e^{  - \Delta \tilde{H}/T }$, 
where $\Delta \tilde{H} = \tilde{H}_1 - \tilde{H}_0$, if $\tilde{H}_1 > \tilde{H}_0$. For simplicity, we assume that 
$T$ is the same as in step 2.

\item Continue with the next timestep. 
\end{enumerate}

The parameter $n$ controls the {\em relative update rate} between opinions and links. 
The relative frequency of opinion updates versus link updates is $nN$. 
Depending on the choice of $n$, which can range from zero to infinity, (and depending on the initial conditions), 
the opinions may or may not be given enough time to 
converge towards a steady state between link updates; in other words they may or may not have enough time to ``equilibrate''.
In the SI we show the consequences of different choices of $n$. 
In the main body of the paper we set $n= 1$.
Here we are interested in a true co-evolutionary dynamics, which is guaranteed for this choice of $n$ and the range of $N$'s 
considered in the paper. 
Appropriate care needs to be taken when larger systems are studied to ensure that the co-evolution is correctly implemented.

The parameter $T$ is a kind of {\em `social temperature'}  that 
characterises the average volatility of individuals in a society \cite{Bahr}. 
The higher $T$ is, the more volatile on average an individual is. 
This means that he or she is more likely to update his/her opinion and social ties, 
regardless of which flips reduce social stress. 
The update rules specified by $p$ and $p'$ are based on the intuition 
that a change that reduces social stress (lower $H$) 
is more favourable than one that  increases it. 
The choice of an exponential function is for convenience only and 
has no particular meaning (as it has in physics). 

\subsection{Social coherence through external influences}

Opinion formation is not a purely endogenous process. It can be influenced strongly by external 
influences, such as religion, nationalism, and so on. Within the proposed framework, such influences 
can be included with additional terms in the $H$ function.
We propose to study a term that discourages people from maintaining hostile links.  
This could be the message of an exogenous religious or moral norm (``love all the others''), 
or some nationalist propaganda that suggests that people of the same nation should 
be unconditionally friendly to one another. 
To this end, whenever we want to model exogenous pro-social pressure, 
we  add a third term, $ (h/2) \sum_{(i, j)} (1-J_{ij})$, where $h > 0$, to  Eq. (\ref{eq:model}). 
Clearly, this term will suppress negative links in the society. 

\subsection{Characterising modes of collective behaviour---order parameters for social fragmentation}

To characterise the degree of  social cohesion or fragmentation
we have to define appropriate quantities that we call {\em order parameters}. 
In the theory of phase transitions \cite{Landau}, order parameters signal regime shifts from one  phase into another. 
To quantify the degree of social fragmentation we use the following measures:

\subsubsection{Size distribution of echo chambers}

A clear signal for social fragmentation is the distribution of cluster sizes. 
In a fragmented society there exists a large number of small groups of individuals that cooperate within their group but are 
hostile towards other groups. We detect these clusters by minimising the number of positive 
relations between them and that of negative links within them \cite{Doreian1996, Traag2009, Esmailian}.
By doing that, most of the negative links will be found between the clusters. 
Further,  in agreement with the notion of echo chambers in the literature, 
from the detected ``positive'' clusters, we select those  that consist of only like-minded agents and 
identify them as echo chambers. 
The size of an echo chamber is thus given by the number of such nodes, 
and is denoted by  $S(\mathcal{E})$, where $\mathcal{E}$ denotes  the chamber.   

\subsubsection{A measure for polarisation, $f$}

We introduce a simple network variable, $f$, to measure the level of social balance in the society. 
It is defined as  the difference of the fractions of balanced and unbalanced triangles  in the network: 
 \begin{equation}
 f = \frac{n_{+}  - n_{-}}{n_{+} + n_{-}} \quad ,
 \label{f}
 \end{equation}
where  $n_{+}$ and $n_{-}$ are the  number of balanced and unbalanced triangles, respectively.   
$f= 1$ means that all triangles are balanced, $f < 1$ signals that unbalanced triangles are present. 
Even though $f$ could be negative,  this situation is never observed in simulations. 
This is in agreement with both Heider's intuition and the empirical evidence obtained in real social networks, 
where the value of $f$ is typically above $0.7$ \cite{Leskovec}. 
The case $f \to 0$ corresponds to an equal number of balanced and unbalanced triangles. 
From Harary's result\footnote{A network is balanced if it  consists of only balanced  cycles 
(triangles are a special case of cycle of length $3$). His theorem states  that a signed graph is balanced 
if and only if the set of nodes can be partitioned into two disjoint subsets (one of which may be empty), 
such that all links between nodes of the same subset are positive, and all links between nodes of the 
different subsets are negative.} \cite{Harary},  
it follows that if the network can be partitioned  into strictly positive clusters, within which \emph{all} links are positive 
and between which links are \emph{exclusively} negative, then $f = 1$.  
In reverse, the case, $f = 1$, is not sufficient to imply such a partition for sparse networks; 
however, high values of $f$ ($f \to 1$)  generally correspond to a clustering that is close to this partition.

\subsubsection{A measure for group homogeneity,  $m_g$}

We need a  quantity to characterise how opinions are  distributed within groups. 
If a society fragments, it re-organises into sub-communities (clusters)  
of mutually befriended individuals who would be expected to hold similar opinions. 
We can measure the average level of opinion homogeneity within a group by
\begin{equation}
m_g = \left\langle  \frac{1}{S(\mathcal{C}_k)} \left|\sum_{i \in  \mathcal{C}_k} s_i\right| \right\rangle_{\mathcal{C}_k}    \quad ,
\end{equation}
where $\mathcal{C}_k$ denotes the $k$-th  positive cluster found by the community detection method \cite{Doreian1996, Traag2009, Esmailian}. 
The average, $\langle . \rangle_{\mathcal{C}_k}$, is taken over all the detected clusters.
By definition, $m_g$ is the average of the absolute values of the local (binary) opinions over all groups
so that $m_g \in [0, 1]$.  $m_g=1$ if and only if all clusters are composed of like-minded individuals only. 
However, opinions may be different between individuals belonging to different clusters. On the contrary, 
$m_g = 0$ corresponds to a totally cohesive society that consists of either one or many groups of 
befriended individuals but there is no opinion that dominates in any one of these groups. 
Intermediate values of $m_g \in (0, 1)$ signal that within  a group opinions vary   
and there is no consensus among its members. 

\subsubsection{A measure for opinion diversity,  $m$}

As a simple measure for the opinion diversity across groups we compute the 
overall opinion of the society 
\begin{equation}
m =  \frac{1}{N} \left|\sum_i^N  s_i   \right|  \quad .
\end{equation}
By definition, $m \in [0, 1]$. The lower $m$ is, the more diverse opinions are. 
Opinions are aligned across society if $m \rightarrow 1$. 
This measure can also serve as a probe of how fast opinions can converge to a consensus. 
This is important because in real social contexts one can change opinions and friends (or enemies) 
within a limited lifetime. Therefore, convergence times do matter and must be studied in detail. 
The time required for the system to equilibrate from different initial conditions may vary strongly.

\section{Results}
\label{sec:results}

We simulate the model given in Eq. (\ref{eq:model}) for the parameter choices of $N=400$, and $g=1$.
We first discuss the phase diagram of the model and its consequences. 

\subsection{Phase diagrams}

\begin{figure}[tb]
\begin{center}
\includegraphics[scale=0.3]{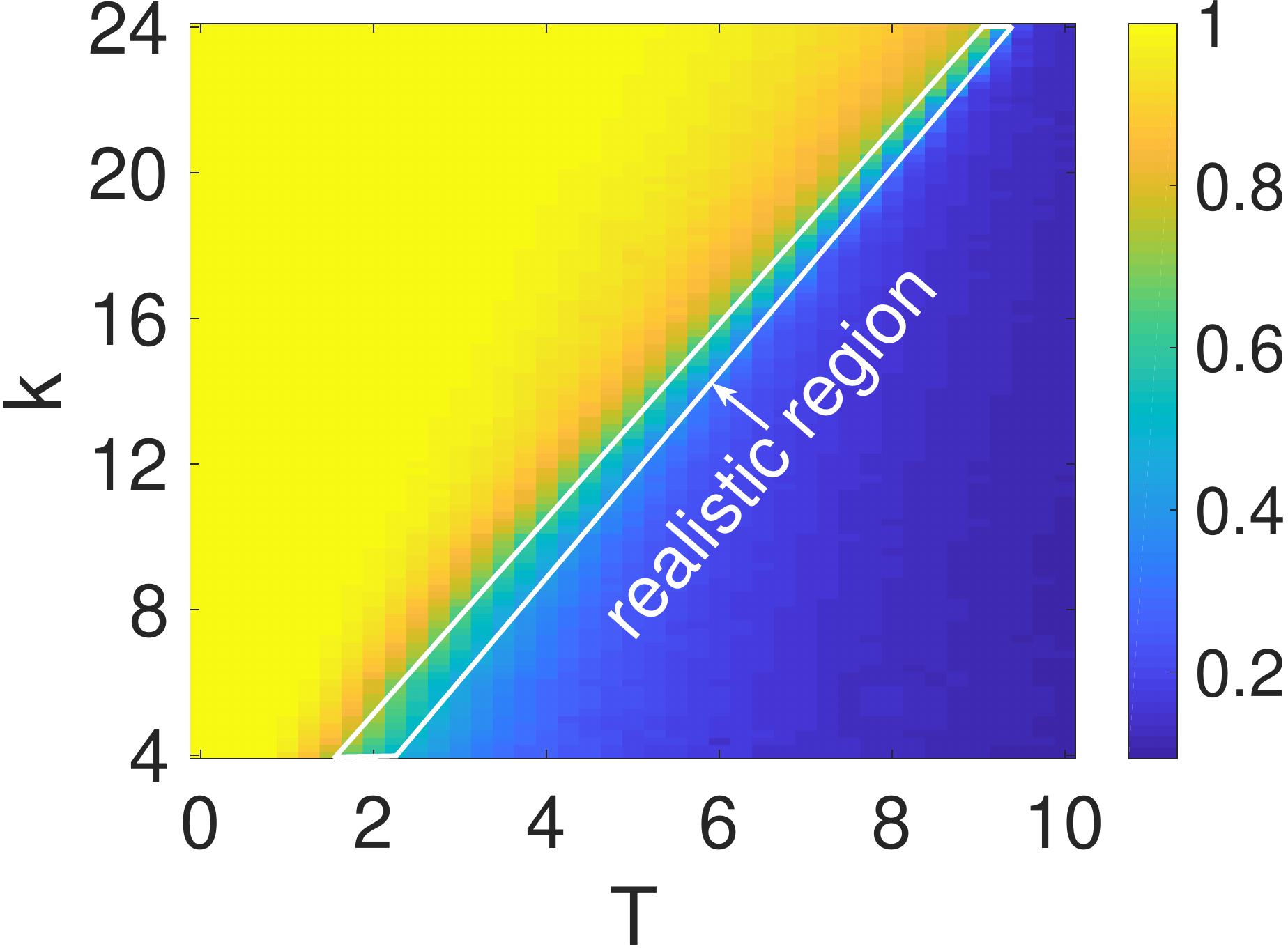}
\end{center}
\caption{
Phase diagram of the stochastic co-evolutionary model with social balance. 
The balance level { $f$} is shown as a function of the average network degree, $k$, and  social temperature, $T$. 
A clear phase separation line is visible. Below it, for low values of connectivity and high $T$, there exits a socially coherent phase (blue), 
above the line there is a phase of social fragmentation (yellow).  
Empirically reasonable values of $f$ around $0.7$ are indicated with a white box.  
Results were obtained for regular lattices ($\epsilon=0$),  $g =1$, $N=400$ and  
are averaged over 500 realisations. Random initial conditions in links and opinions.  
}
    \label{fig:phasediag}
\end{figure}

The central result of this paper is shown in Fig. \ref{fig:phasediag} that shows $f$ (in colour code) as a function 
of the average connectivity, $k$, and social temperature, $T$. 
There is a clear separation line at which  the society transitions from a 
well mixed situation with $f\sim 0.1$ (blue) to a fragmented one, characterised by $f\sim 1$ (yellow). 
In the yellow region, the emergent networks are strongly balanced and opinion clusters exist.  
These polarised clusters disappear and opinions become randomly distributed amongst agents in the dark blue region, 
where there are as many balanced as unbalanced triangles.
Note, that values of $f\sim 0$ are unrealistic. 
Real societies are balanced and  show empirical values in a range around $f\sim 0.7$ \cite{Leskovec,Szell}. 
We indicate the realistic region with a white box. 
Assuming that a given society is found somewhere  in the realistic region, say at a fixed $T$, it only takes 
a small increase of social connectivity, $k$, for the society to be pushed into the fragmented filter bubble phase. 
In recent years the average connectivity has certainly increased in societies, 
making it easier for them to transition into the fragmented regime.
The result in Fig. \ref{fig:phasediag} is obtained for a regular lattice ($\epsilon=0$). 
We confirmed that the existence of the separation line also holds for small world network topologies; see also below. 

\subsection{Size distribution of echo chambers}
 
\begin{figure}[tb]
\begin{center}
\includegraphics[scale=0.24]{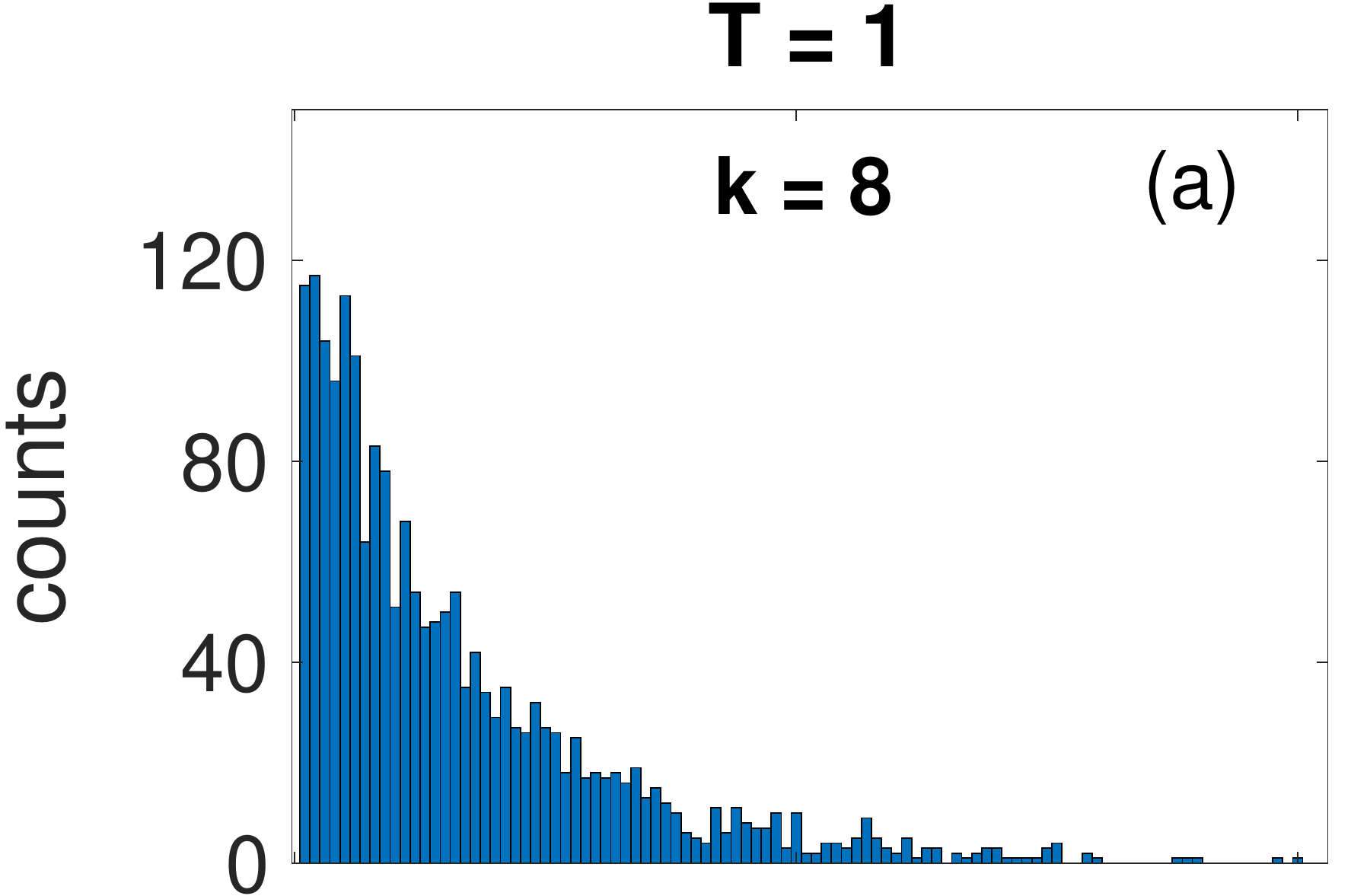}
\includegraphics[scale=0.24]{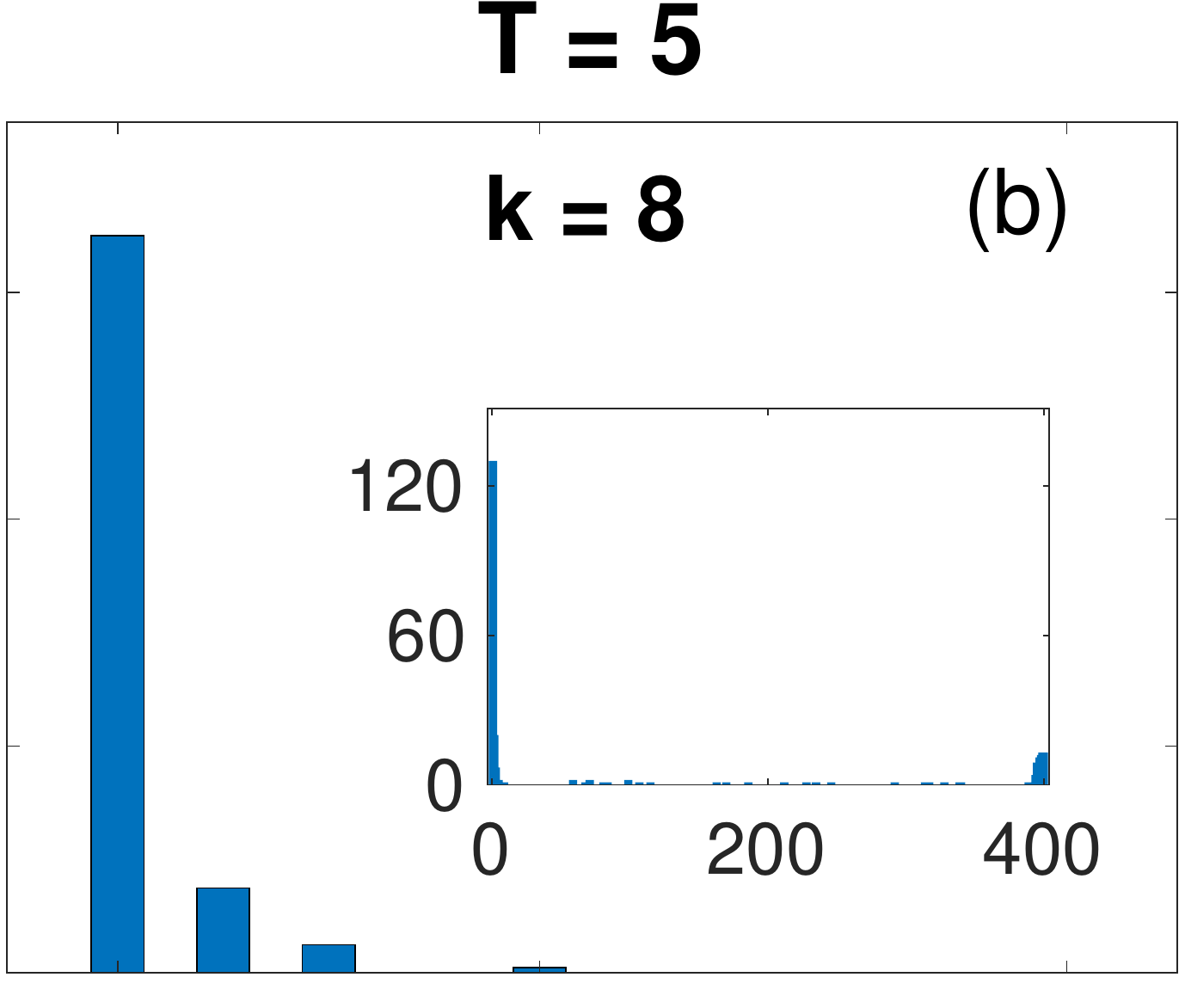}

\includegraphics[scale=0.24]{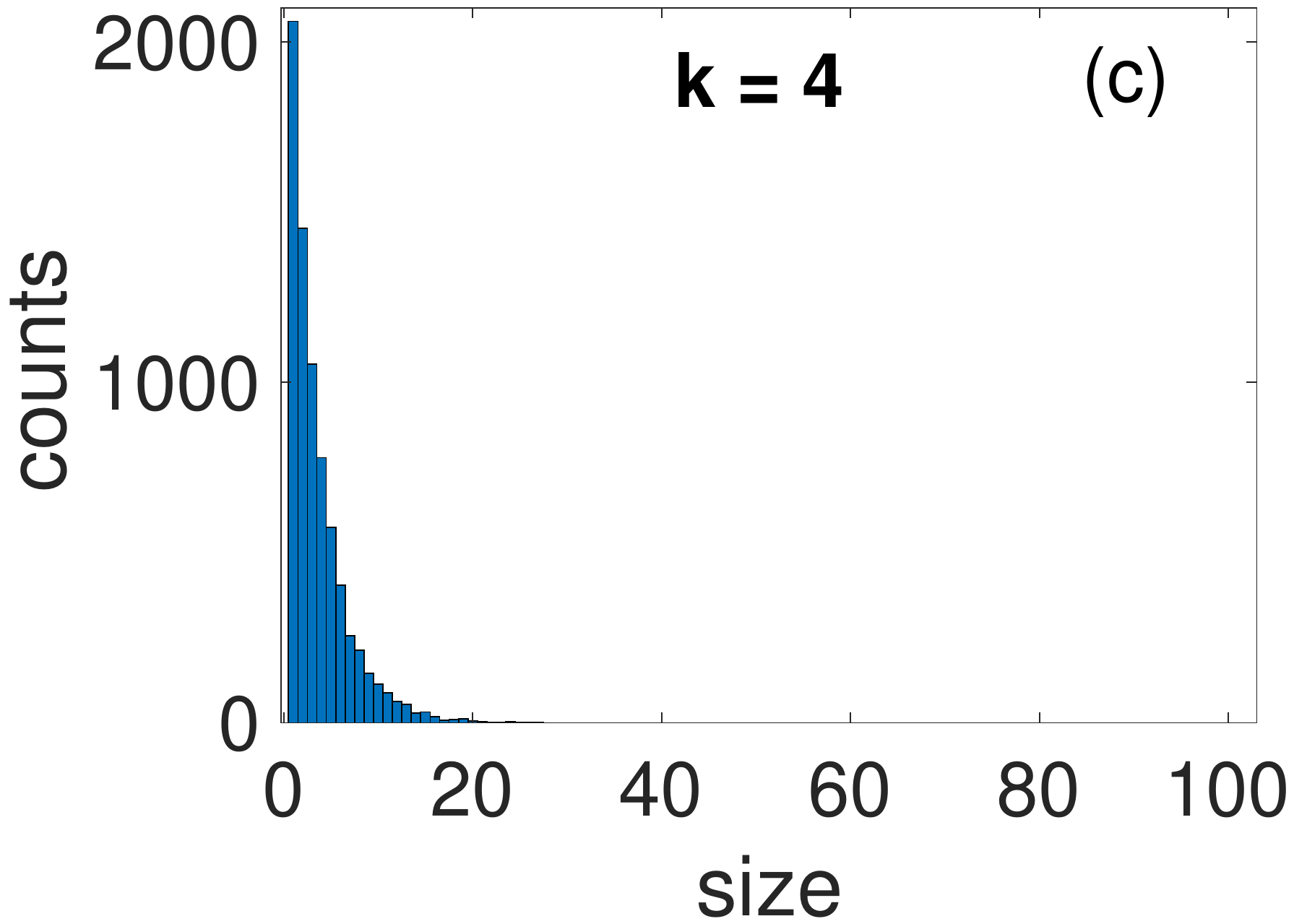}
\includegraphics[scale=0.24]{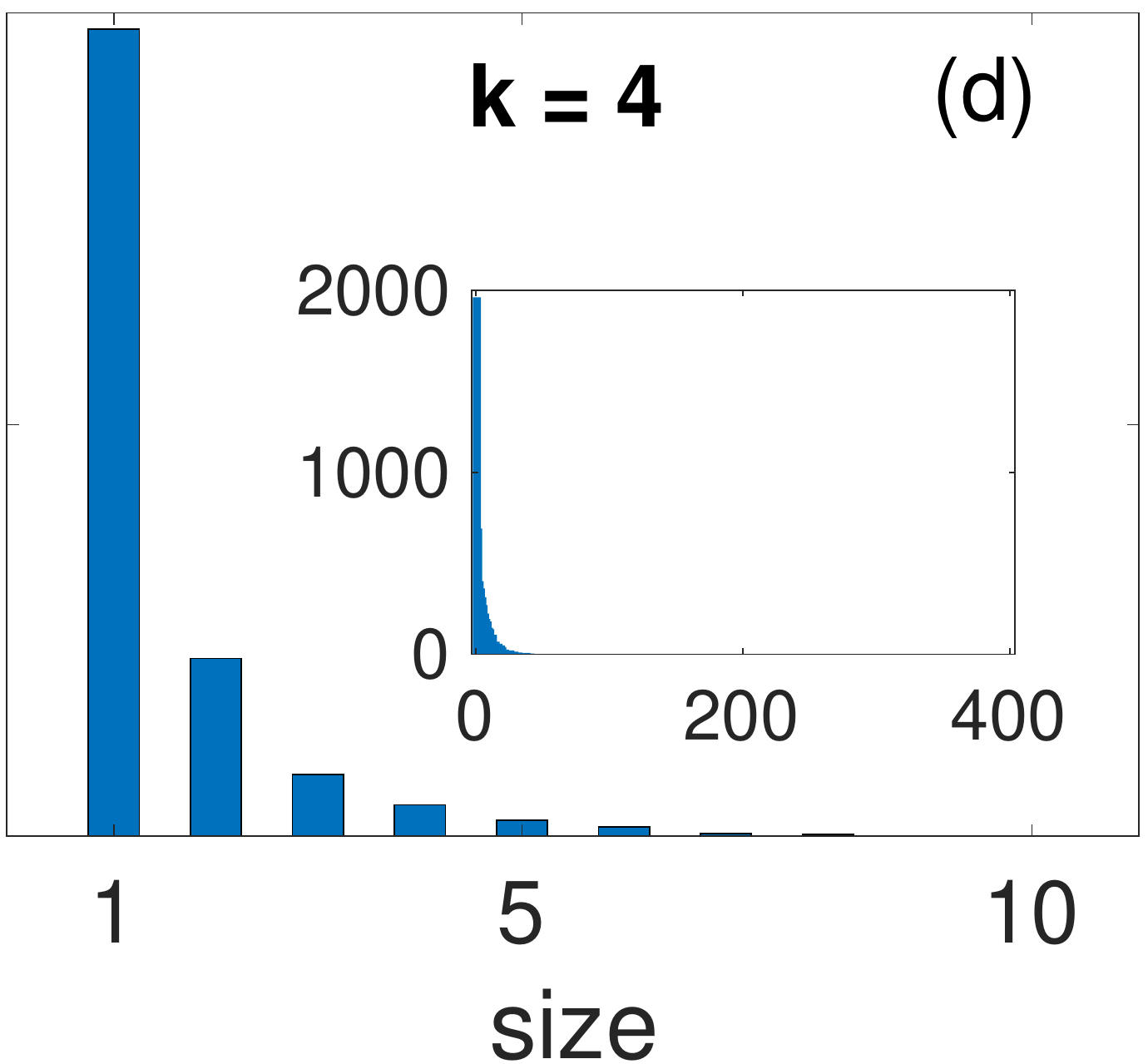}
\end{center} 
\caption{Distributions of echo chamber sizes  as a function of average connectivity, $k$, and social temperature, $T$.
Echo chambers are defined as groups of friendly agents who hold the same opinion.
The left column represents the situation in the fragmented phase (low temperature $T=1$). 
The right column is in the cohesive phase. The upper panels show an average connectivity of $k=8$, the lower ones $k=4$. 
Clearly, in the fragmented phase there appear significant groups of all sizes that are characterised 
by uniform opinions and positive relations within, and hostile relations towards others.   
The insets show the size distribution of  the detected ``positive'' clusters, ${\mathcal{C}_k}$. 
These are groups of cooperating individuals, where any two members of the same group can be 
connected by a path consisting of positive links only. In the cohesive phase there exist positive clusters of 
maximal sizes ($S({\mathcal{C}_k}) = 400$), meaning that the whole society can cooperate  
despite a diversity in opinions.
Same model parameters as in previous figure. 
} 
   \label{fig:clusters}
\end{figure}

We show the echo chamber size distribution for various values of $T$ and $k$ in Fig. \ref{fig:clusters}. 
The left column shows the situation for small social temperature, $T=1$, the right column shows $T=5$. 
The upper panels show a high average connectivity, $k=8$, while the lower ones correspond to $k=4$ neighbours. 
The left column corresponds to the fragmented phase, the right column to the cohesive phase. 
It is clearly visible that deep in the fragmented phase there is a broad distribution of echo chamber sizes,
spreading to sizes of about 100 for $k=8$ 
and to sizes of about 20 for $k=4$. 
In the right column we observe sharply peaked distributions with maximum cluster sizes of about 2-3, 
meaning that there is no large cluster of unique opinion forming. This corresponds to a society where 
different opinions co-exist. 
The insets show  the size distribution of  the ``positive'' clusters ${\mathcal{C}_k}$ found by the 
community detection method. Note that in (b) there is a small peak at 400, which is the maximal size of a cluster. 
This indicates the possibility of global cooperation of the whole society in the cohesive phase even if opinions are diverse. 

\subsection{Robustness}
To find out if results are robust with respect to changes of parameters, 
we perform a series of robustness checks. 

\begin{figure*}[t]
\begin{center}
\includegraphics[scale=0.3]{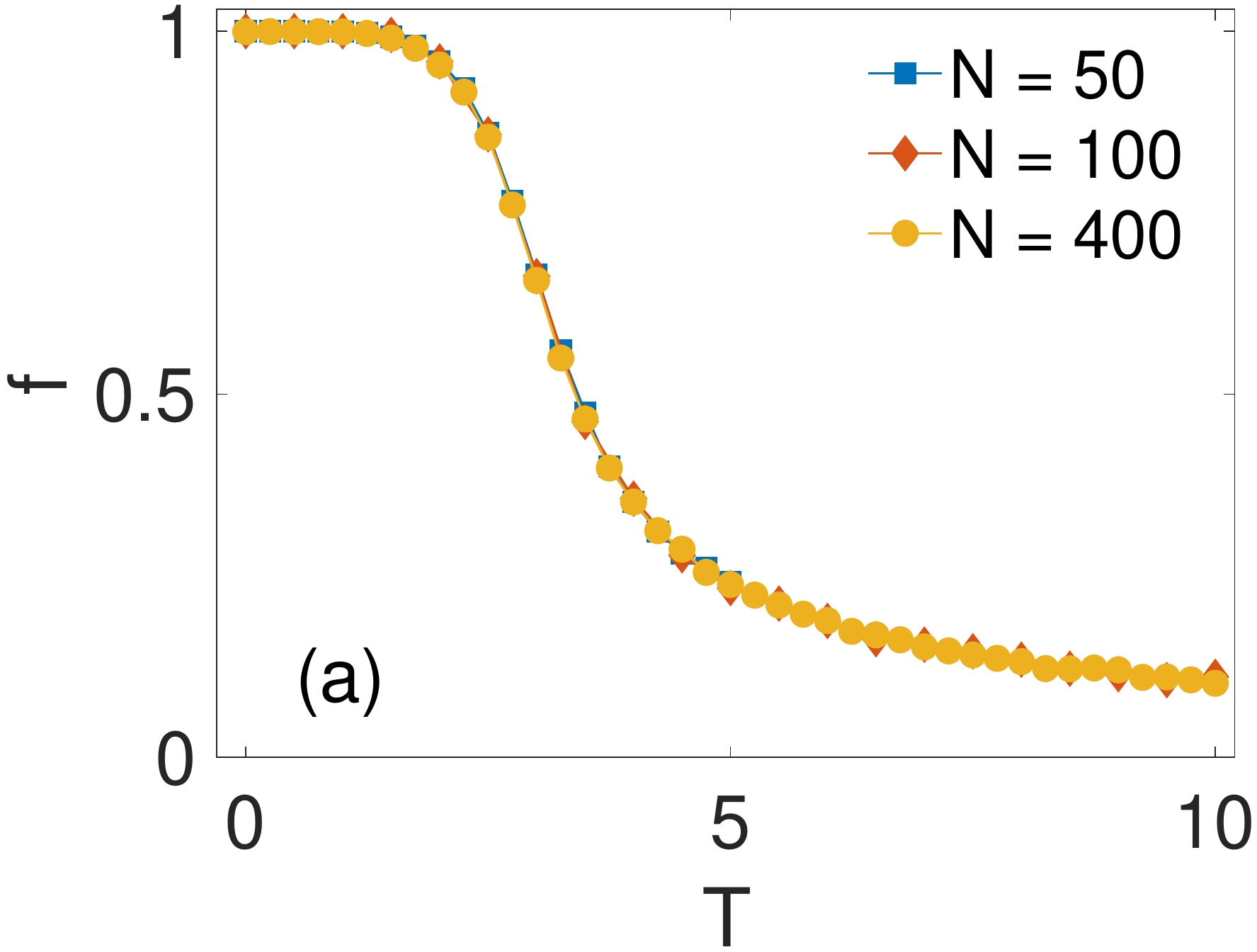}
\includegraphics[scale=0.3]{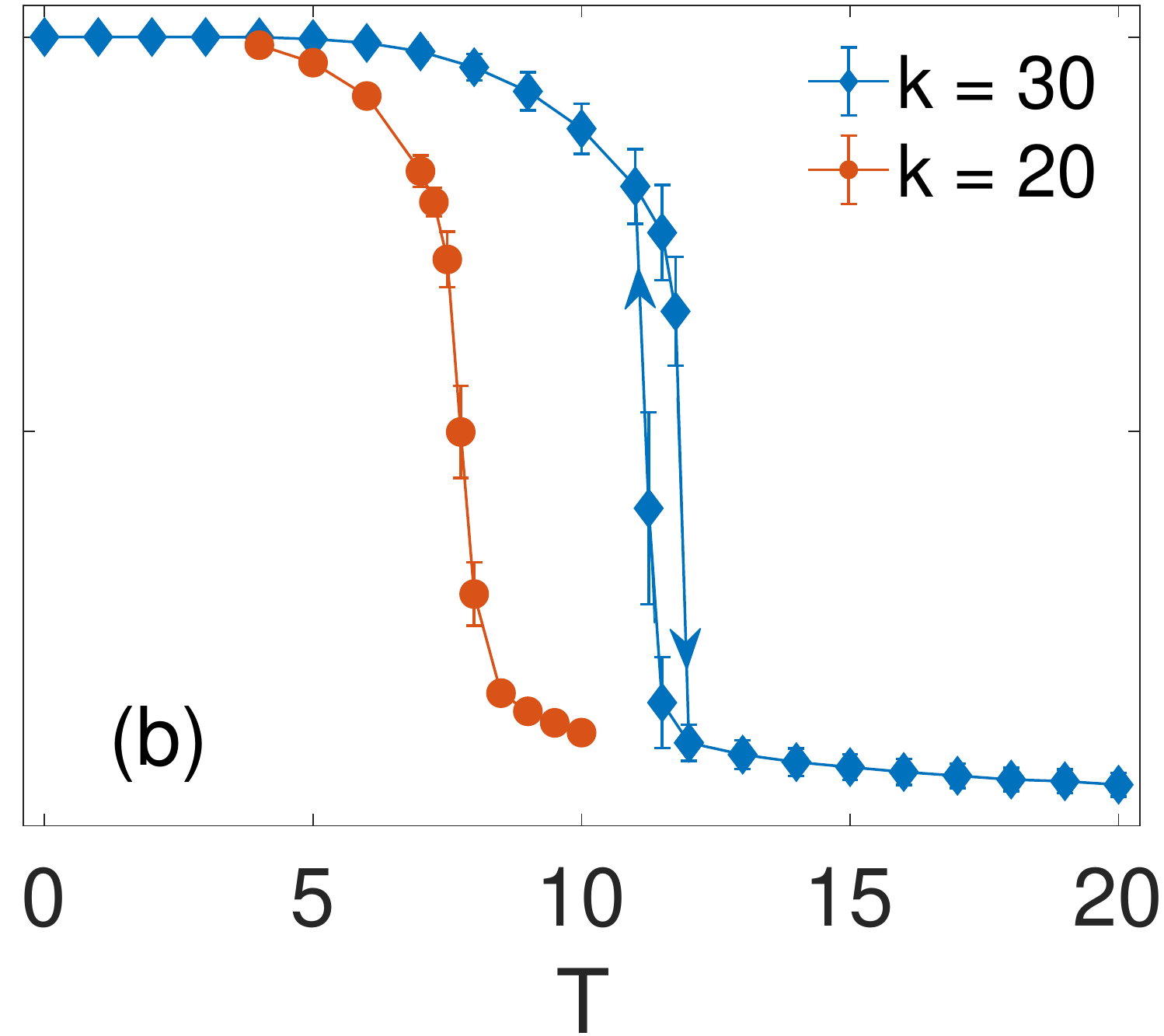}
\includegraphics[scale=0.3]{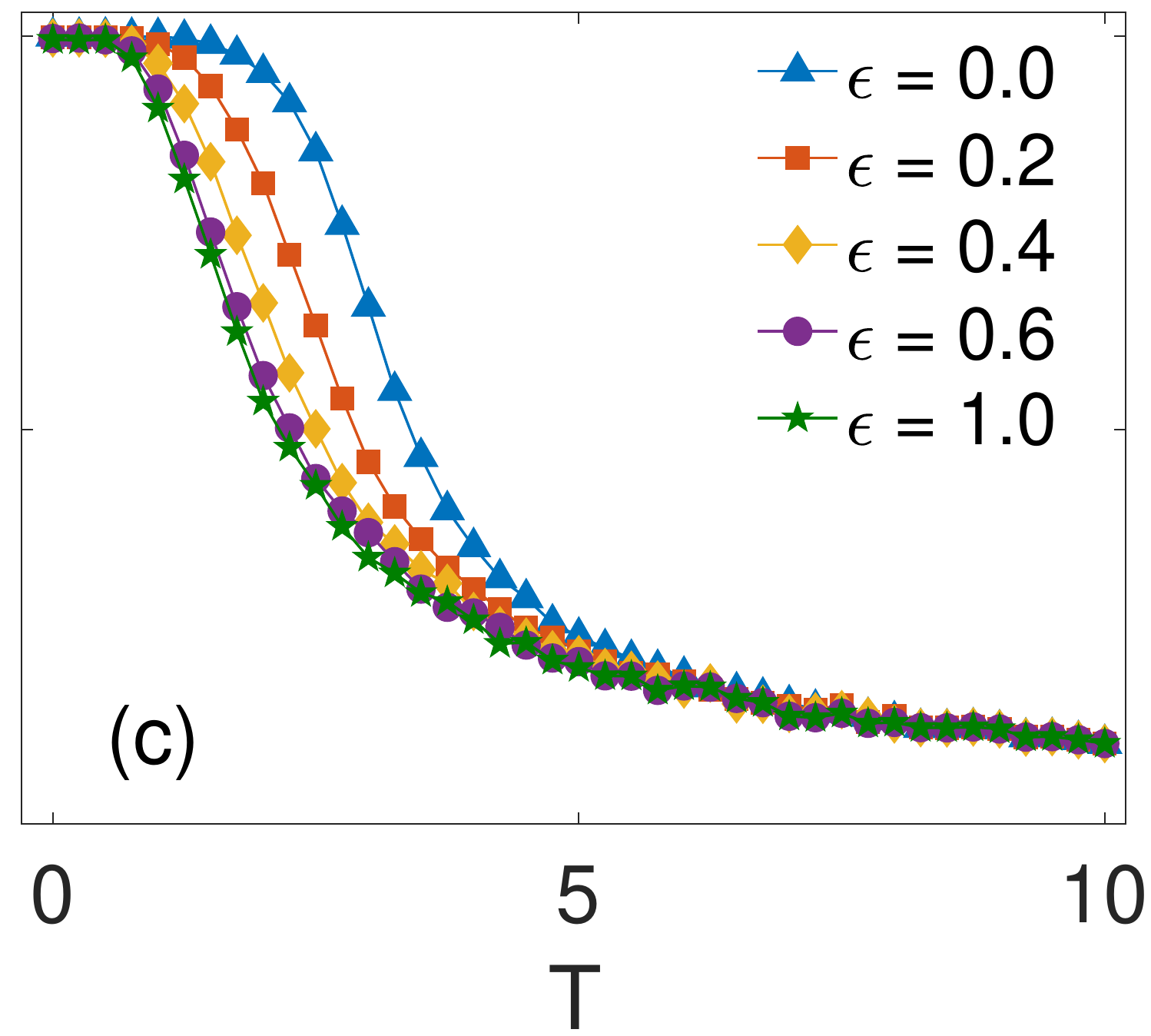}
\end{center} 
\caption{Robustness of the results.
(a) Size dependence. Section of the phase diagram for $k=8$ for various sizes of society $N = 50, 100, 400$.
No size dependence in the phase diagram is visible. 
(b) Higher average connectivity pushes the phase transition towards higher critical temperatures. 
A discontinuous transition of $f$ is observable that shows a hysteresis effect. 
Note that it is especially pronounced for large connectivities. 
The existence of this hysteresis could indicate a potential handle to avoid fragmentation, see discussion. 
$N=400$, results are averaged over $100$  independent realisations of the model.
(c) Change of the phase transition for a small-world network structure with $\epsilon = 0, 0.2, 0.4, 0.6, 1$. 
$k =8$, $N = 400$, $g = 1$. Results averaged over 200 realisations for every $\epsilon$.
}
\label{fig:f}
\end{figure*}

We first test the dependence on the size $N$ of the society.
For a fixed value of $k=8$ we show a section of the phase diagram in Fig. \ref{fig:f} (a) 
for various system sizes, $N=50, 100, 400$ on regular networks. 
Clearly, there is no visible size dependence. It can be safely assumed 
that this will also hold when taking $N\to\infty$\footnote{In this case 
the relative update ratio of opinions to links has to be modified appropriately.}. 
This result is not unexpected since we keep the connectivity smaller than $N$. 

In Fig. \ref{fig:f} (b) we show  the effect of the average connectivity on the results, where 
we fix $N= 400$ and compute $f$ for various values of $k$.
As we already noted in Fig. \ref{fig:phasediag} with increasing $k$ the phase diagram is shifted 
towards the fragmented phase (yellow region in the phase diagram). 
The transition appears to be discontinuous ({\em first-order}), meaning that $f$ jumps as a function of the temperature variable.  
Fig. \ref{fig:f} (b) also demonstrates a hysteresis effect (visible for $k=30$), which  often accompanies first order transitions.
This can be understood in the following way: If in Fig. \ref{fig:f} (b) we increase $T$, $f$ starts to gradually decrease, and then drops 
rapidly to much lower values. If at that point one would start decreasing $T$, $f$ would not immediately jump up to previous levels, but remain low until at a lower $T$ it would finally jump upward again.  See arrows in the figure. 

To test if the particular network structure has an influence on the results, we computed the 
phase diagrams with small-world networks \cite{WattsStrogatz}. 
The small-world parameter, $\epsilon$, controls the probability to re-connect a link from any node to any other node. 
Here we rewire the  connections in such a way that the network  remains connected and does not 
dissociate into different components. 
Note that $\epsilon=0$ means a regular network, $\epsilon=1$ corresponds to a random graph.
Figure \ref{fig:f} (c) shows the result. The transition line is shifted towards the left, i.e., 
the critical temperature decreases with increasing $\epsilon$. 
This fact can be understood as a consequence of having less triangles in the networks 
that are obtained with a larger value of $\epsilon$; see SI \ref{fig:trianglesinWS}. 
  
Finally, we check what happens if we lower the coupling strength of the Heider term in Eq. \ref{eq:model}. 
When we take $g=0.01$, we observe a shift of the  phase transition line to the left and the 
dependence of the transition on the connectivity, $k$, becomes negligible; see Fig. \ref{fig:g001} in the SI. 
Obviously, for the case $g\to0$ where the Heider term vanishes, there will be no more dependence on $k$. 
The pronounced fragmentation transition at high interconnectedness is hence a direct consequence of social balance.

\subsection{The role of external influences}

\begin{figure}[tb]
\begin{center}
\includegraphics[scale=0.23]{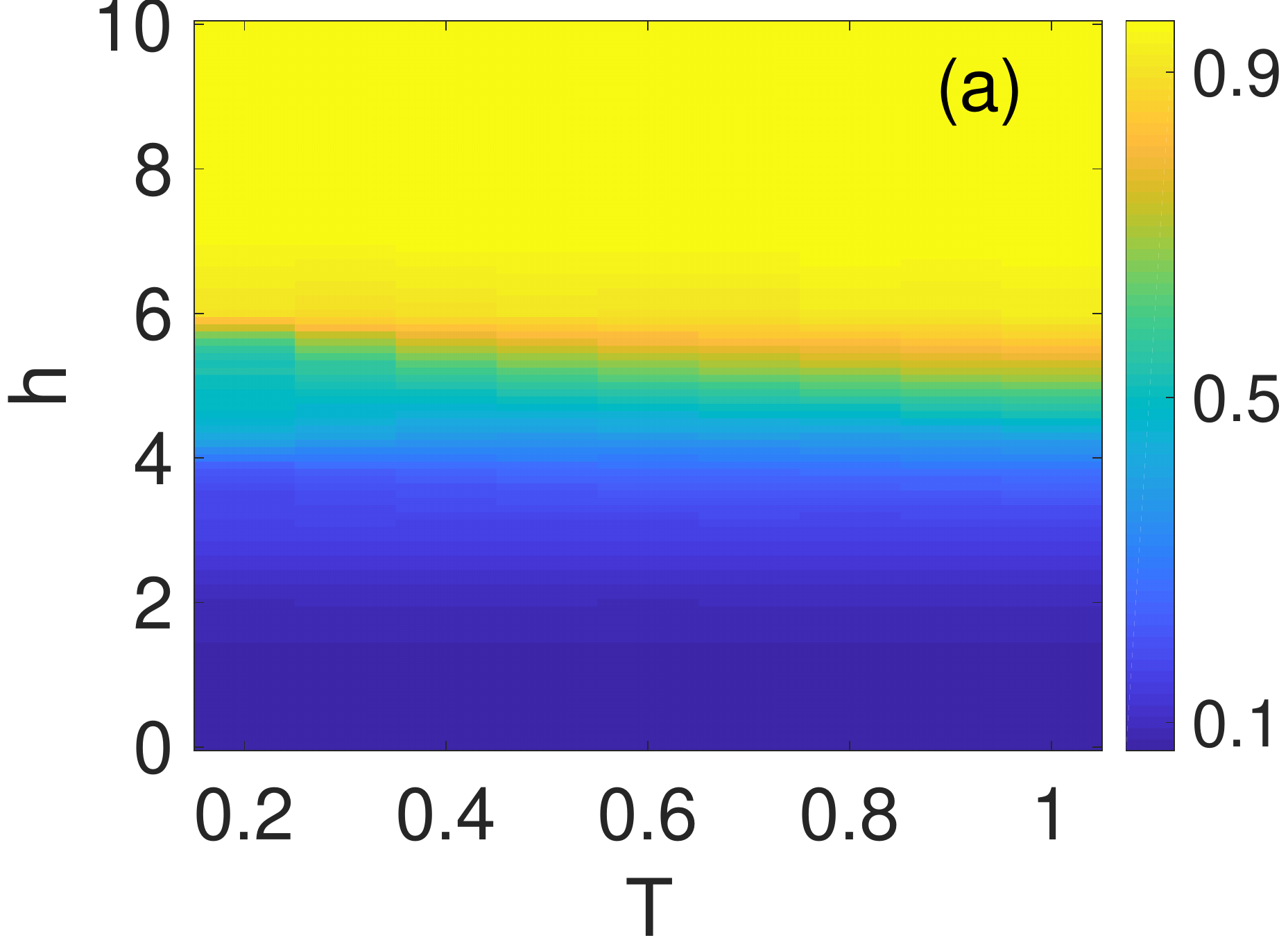}
\includegraphics[scale=0.23]{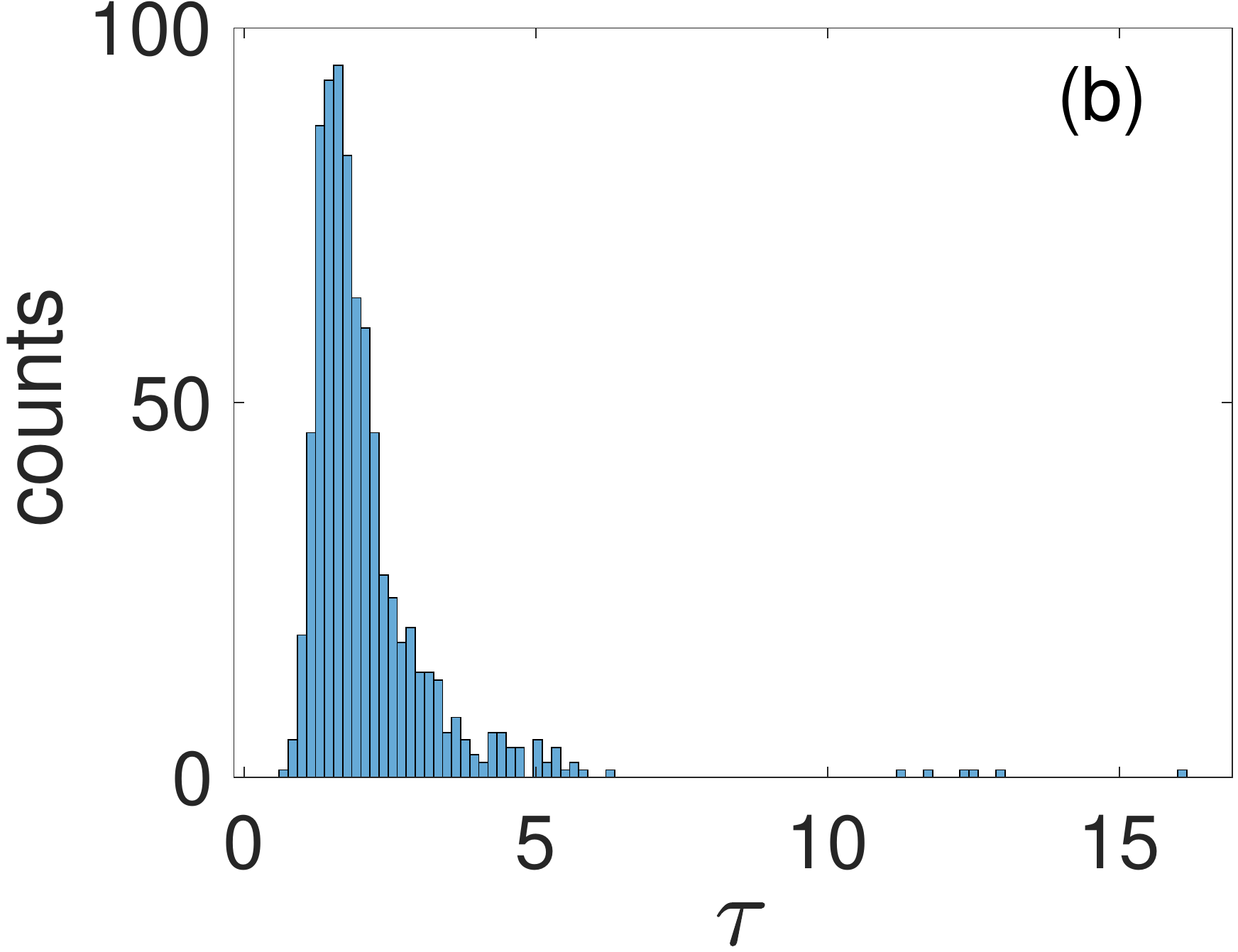}
\end{center}
\caption{ 
(a) Opinion diversity (colour), $|m|$, as a function of social temperature, $T$,  and the external influence parameter, $h$.   
The blue region indicates the case where  opinion clusters exist. 
In the yellow region global consensus is the unique attractor of the dynamics. 
Note the change of meaning of yellow and blue with respect to Fig. \ref{fig:phasediag}. 
The two phases of high and low $|m|$ are separated by a critical line $h_c = h_c(T)$.  
Note that the formation of global alignment of opinions takes very long ($O(10^4)$  Monte Carlo iterations).  
$N = 200$, $k = 10$, $g = 1$, $\epsilon=0$, results averaged over $3,200$ realisations. 
(b) Distribution of convergence times, $\tau$, at $T = 1$. Results were obtained for regular networks 
$\epsilon =0$, $g = 1$, $N = 200$, $k = 10$ and  are averaged over 800 realisations. 
$\tau$ is measured in the unit of $kN$ timesteps. Its mean is $\langle \tau \rangle \simeq 2.11 \times kN$, 
its variance is $\sigma^2_{\tau} \simeq 1.3 \times kN$.
}
    \label{fig:external}
\end{figure}

In Fig. \ref{fig:external} (a) we show  the effect of the external influence, $h$, 
designed to suppress negative links in the society, on the fragmented phase. It does what it is expected to do. 
Note that the terms $h$ and $g$ may compete with each other: if $h$ promotes the flip of a negative link this could 
result in more unbalanced triangles, meaning that it works against the effect of $g$. 
As a consequence of this competition, a low value of $h$ can only remove a small fraction of negative links 
and a fragmented society emerges, similar to the case without the external influence. 
Only beyond a critical threshold, $h_c$, can most of  the negative links be eliminated and global consensus be 
reached. See SI \ref{fig:path} for an illustration of this phenomenon for a simple network of $N = 3$ nodes.  
Since there is no transition in $f$ (it remains close to $1$), we use $m$ to characterise the change in the 
final state of the society under the effect of the external influence.

 \subsection{A note on time scales}

In Fig. \ref{fig:external} (b) we analyse the times, $\tau$, that are necessary for the order parameters to converge 
 to their stationary values. This is essential  to check since convergence times in this type of system can be 
exceedingly and unrealistically long. 
$\tau$ is the time required for the system to equilibrate at low social temperature.  
We observe in Fig. \ref{fig:external} (b) that on average $\tau$ is of the order of $k N$ timesteps. 
Given that the number of links is $kN/2$, the distribution of $\tau$ with a mean $2.11 \times kN$ means   
that the network updates about 4 times on average before reaching equilibrium. 
However, for a particular run, $\tau$ may vary substantially, depending also on the initial conditions. 
Typically, the steady state can be reached faster if  the initial fraction of positive links is above $1/2$. 
The variability becomes  more pronounced in the presence of external influence, $h$. 
At sufficiently low temperature, the convergence times can become very long due to  the 
existence of many local minima, so-called ``jammed states'', in the energy landscape \cite{Antal2005, Marvel}.   
The  social stress in a jammed state is  not larger than that in any of its neighbouring states, 
which can be reached from this state by a single spin or link flip.  
Evolving on such a ``rugged landscape'', the system is very likely to get trapped in local minima.  
The global minimum of the social tension, $H$, hence may even become unobservable 
during simulation time.

\section{Summary and discussion}
\label{sec:discussion}

We proposed a model that captures five key elements of human societies: 
(i) {\em Agency.} Humans make their decisions individually. 
(ii) {\em Social context---social networks.} Individuals are constantly influenced by opinions and actions of others in their 
social neighbourhood, or by other external influences. 
(iii) {\em Stochasticity.} Individuals are not fully rational and take random decisions from time to time, that do not maximise 
certain objective- or utility functions. 
(iv) {\em Co-evolution.} Individuals update their opinions as well as their social links. 
Most of these updates tend to avoid social tension.  
(v) {\em Social balance.} Social networks show robust overall structures of 
positive and negative social links. They follow robust patterns of social balance. 

We implemented a simple model that captures these five building blocks in a stochastic manner in the framework of  
a Hamiltonian approach. The focus of the model rests on the notion that humans tend to update opinions and social links, so as 
to reduce social tension. The model exhibits a clear phase diagram, i.e. it shows at which parameter values tipping points 
occur where a society rapidly changes its microscopic composition and structure. 

The results deliver a very clear and robust message:
A society with the ability of a co-evolutionary dynamics of opinion- and link formation 
must be expected to have a phase diagram as the one presented in Fig. \ref{fig:phasediag}.
This is a direct consequence of the social balance term in the model, that incorporates the 
empirical fact that societies are socially balanced to a high degree. 
The  phase diagram shows the existence of a critical connectivity, $k_c$, between individuals of a  society at a fixed 
social temperature, $T$, that controls the update frequencies of opinions and links. 
Below that connectivity, $k_c$, society is in the cohesive phase, where opinions co-exist. 
Above the critical connectivity, society fragments into clusters of individuals 
who share positive links within the clusters and have negative links between groups. 
Within the clusters, large patches of uniform opinions form, and a strong reinforcement of homophily is observed. 
The existence of a critical connectivity is an extremely robust fact;  
if the connectivity increases above the critical value, society inevitably {\em must} fragment.

The model also gives clear answers to how the fragmented phase can be avoided. 
There are only two ways out: either to lower the connectivity below the critical density, $k_c$,
by reducing the number of interaction partners (social distancing)
or, alternatively, to increase the social temperature, $T$, meaning that people would update 
their opinions (and links) randomly more often. 
There are no other alternatives  within the framework of this model. 
For the case of increasing update rates, however,  the existence of the mentioned hysteresis phenomenon must be taken into account. 
This means that if at a fixed interaction density, $k$ update rates, $T$, are increased, the fragmentation might transition 
rapidly to the mixed opinion phase, at, say $T'$. If then the update rates are again reduced,  fragmentation does not immediately 
return, but might reappear at lower update rates, $T''<T'$.  

With the strength of the Heider term, $g$, and the irregular patterns of the underlying networks 
for $\epsilon > 0$,  the position of the critical lines can be shifted. 
The phase diagram remains robust to changes of the overall size of the society. 
We have seen that under strong exogenous influences, $h$, such as religion or nationalism, there is a possibility of 
transitioning from a fragmented society to a ``utopian'' or fascist one;  
such interventions will force the society towards a global consensus.

The presented model has a number of shortcomings. 
Several essential features of real societies have not been included. 
We strongly simplified the structure of social connectivity. Whereas social systems are multi-layer networks, here we have 
focussed only on a single layer. It remains to be seen how the phase diagrams change under the integration of more than one 
layer of (positive and negative) social interactions. 

We have also made simplifying assumptions about the plasticity of  social networks. In reality 
individuals can not only switch the sign of social links, but also eliminate and establish new links. We have made 
a few exploratory steps in this direction, however decided to keep the  topologies fixed for the sake of identifying the essential 
underlying mechanisms. By allowing for more plasticity in network formation, 
we think that the essence of the model will not be affected much.  

The use of one single binary opinion is minimalistic and unrealistic. 
It would be much more realistic to use multiple opinions such as cultural features in the Axelrod  model \cite{RAxelrod1997}. 
The original dynamics, however, needs to be modified to account for negative links. For example,  two agents connected by a positive link can become more similar after interaction, while those who are hostile to each other should grow further apart in the  space of opinions. It would be interesting to compare  the effect of social balance on the fragmentation  in this case with the one that occurs in the presented model. The key message of our model should remain valid as  the social balance ensures the existence of clusters of positive links, within each of which opinions are driven toward uniformity by  the  reinforcement effect of homophily, regardless of the opinion multiplicity.


The use of the same social temperature for both the opinion and link update, is  not justified {\em a priori} and has been 
applied for the sake of simplicity. To describe situations, in which, either agents' opinions are more frequent to change than their relations, 
or vice versa, we introduced the parameter, $n$. As shown in the SI, within a range of $n$ that ensures a true co-evolutionary dynamics, the results do practically not depend on $n$. Alternatively, a stochastic dynamics with two temperatures, one for opinions and one for links is certainly reasonable.  However, in this generalisation a more complicated non-equilibrium approach is required. 
The structure of the  phase diagram may become richer with long-lived metastable phases. Such a non-equilibrium  approach has been considered recently in \cite{Saeedian2019}, where the network evolution is not driven by Heider's balance, but by another aspect of cognitive dissonance. There fragmentation emerges either as an absorbing steady state of the dynamics or  from an active phase due to fluctuations in systems of finite size. 

Finally, from a technical side, the model employed here is a variation of a spin glass model used in physics.  
With the present choice of model parameters (low connectivity,  networks of finite size, $n = 1$), 
we can not expect to find the complicated phase space structure of a mean-field spin glass \cite{Mezard}.
However, the essence of frustration imposed by the Heider term is clearly the same as in spin glasses. 
A more detailed technical study of the model is going to be published elsewhere.

\section{Final conclusion}
\label{sec:conclusion}
The main conclusion of this paper is that it unambiguously shows that the presence of social balance 
carries the seed to social fragmentation. Fragmentation inevitably occurs in a co-evolutionary society 
if the average interaction density exceeds a critical threshold.  

\section*{Acknowledgements}
\label{sec:acknowledgements}
This work was supported in part by Austrian Science Fund FWF under P P 29252, 
and by the Austrian Science Promotion Agency, FFG project under  857136. 
Simulations were carried out in part  at the Vienna Scientific Cluster. 
We thank Tobias Reisch for technical assistance with the simulations.

\bibliography{Spinglass} 

\newpage

\section{Supplementary Material}

\subsection{Dependence of fragmentation on social balance, $g$}

\begin{figure}[htb]
\begin{center}
\includegraphics[scale=0.3]{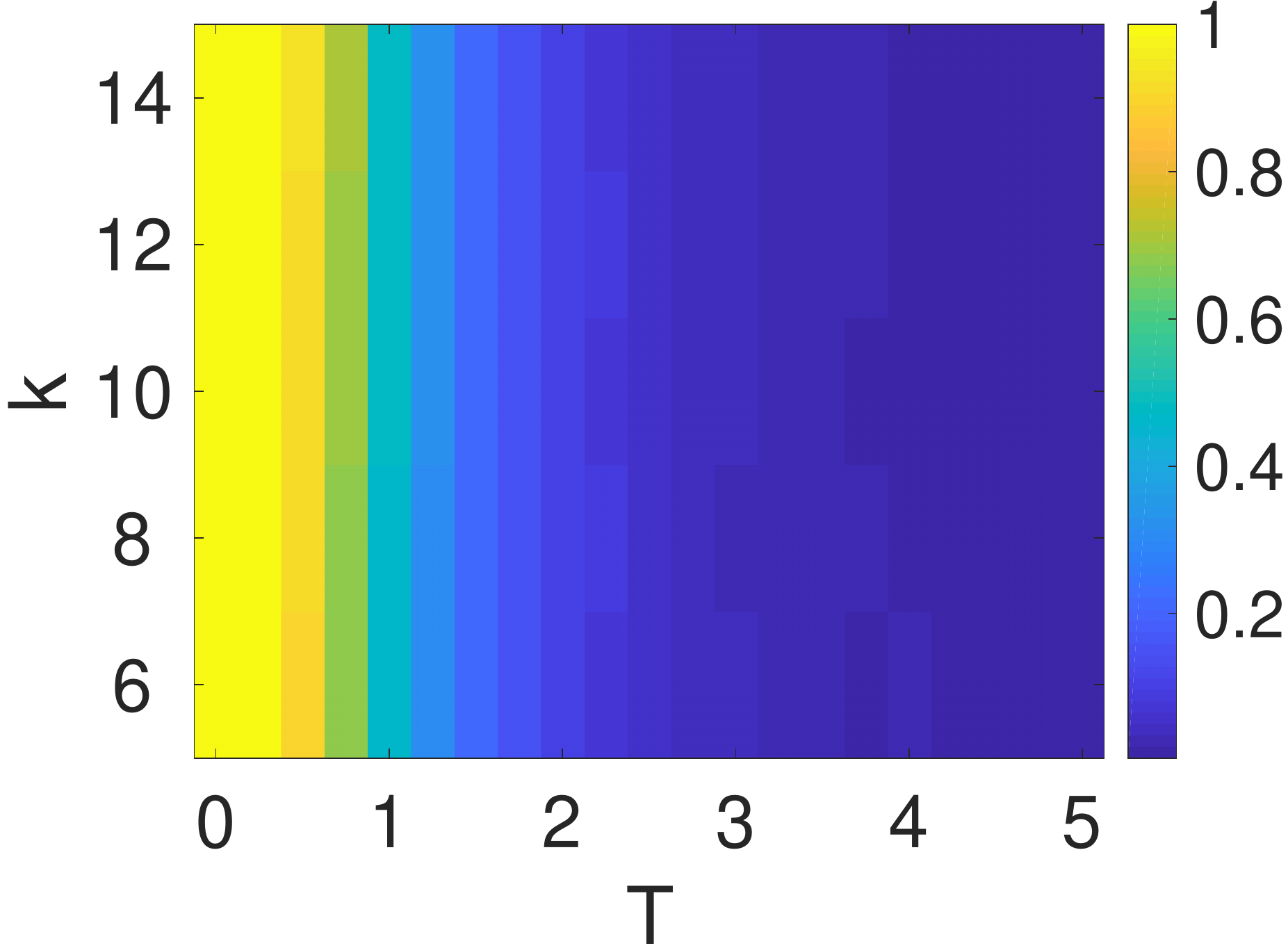}
\end{center}
\caption{Social balance level, $f$, as a function of the average network degree, $k$, and the social temperature, $T$, 
for $g = 0.01$. All other  parameters are the same   as in Fig. \ref{fig:phasediag}. 
Although the fragmented and cohesive phases are separated from each other by a critical line,  
this line becomes vertical, i.e., it does no longer depend on the connectivity. 
The result thus demonstrates the crucial effect of social balance on  the transition of societies between cohesion and fragmentation.
}
    \label{fig:g001}
\end{figure}

\begin{figure}[tb]
\begin{center}
\includegraphics[scale=0.26]{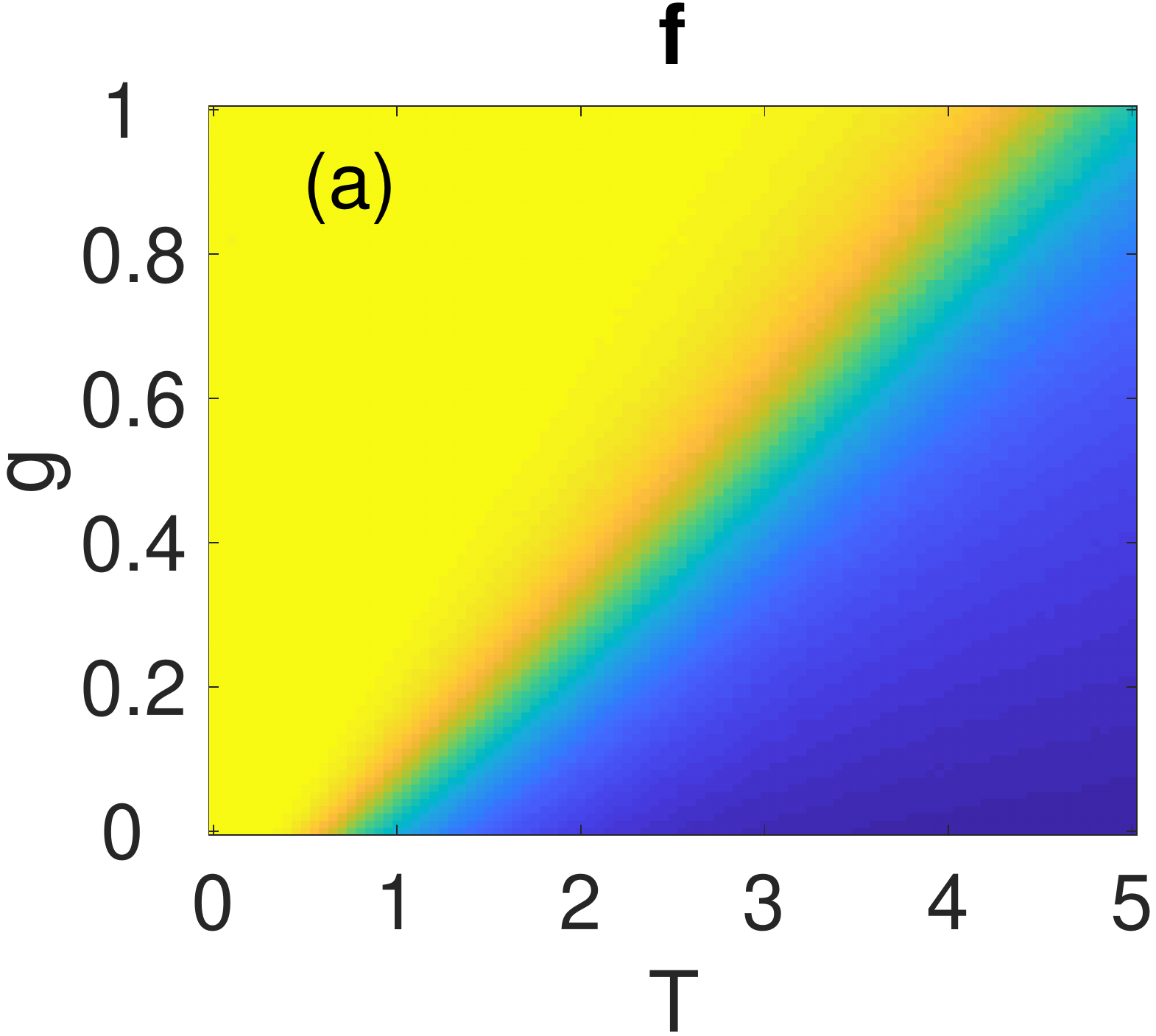}
\includegraphics[scale=0.26]{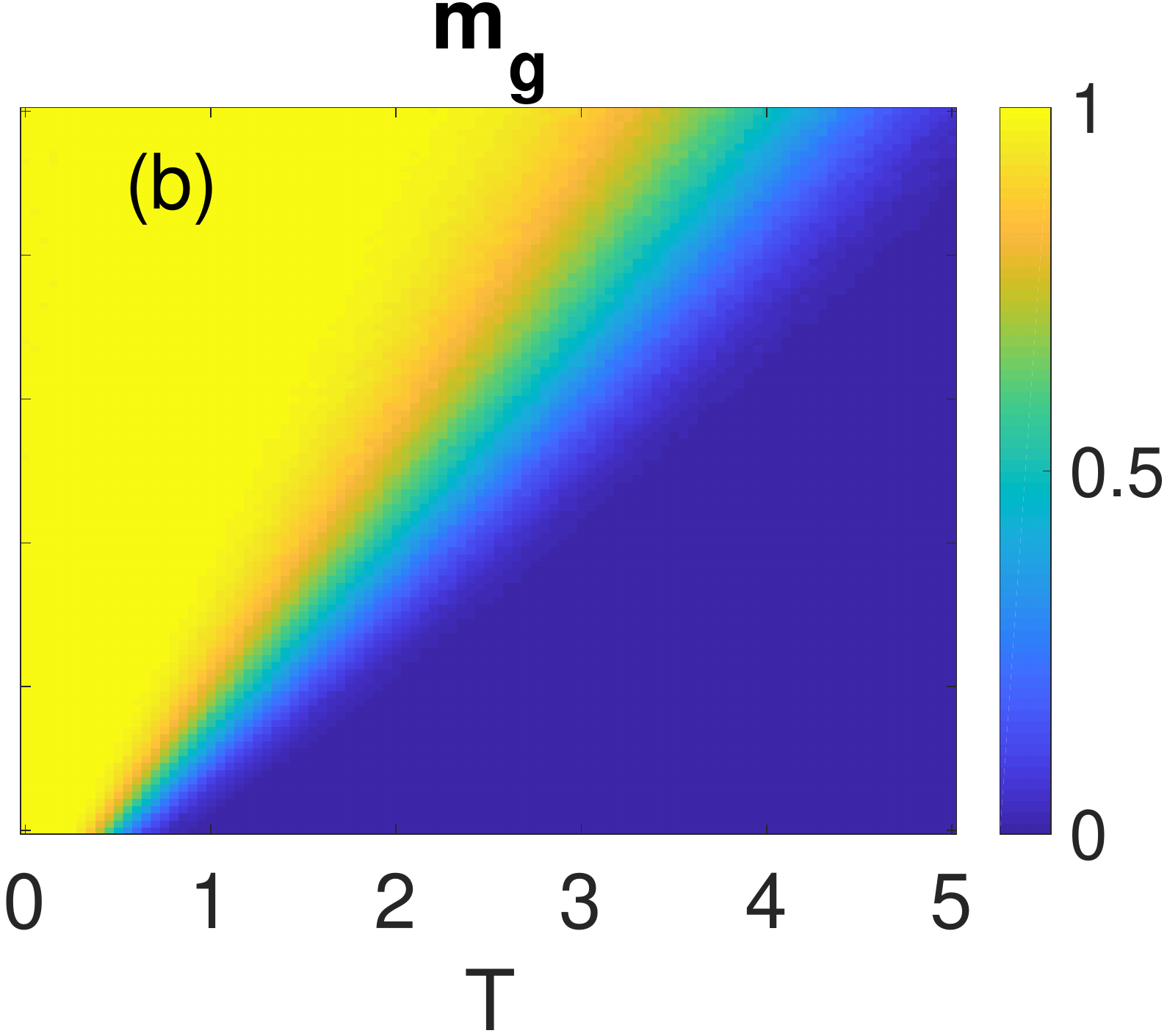}
\end{center}
\caption{
(a) Balance level, $f$, and (b) group homogeneity, $m_g$, both as a function of $T$ and $g$. 
Results averaged over $1000$  realisations for   $N = 10$. Random initial conditions in links and opinions.
The reason for using $m_g$ here is that $m \simeq 0$ in both regions and therefore can't be used to distinguish the two phases. 
}
    \label{fig:phasediag2}
\end{figure}

As shown in Fig. \ref{fig:phasediag}, fragmentation can occur at any level of interconnectedness if the social temperature is low enough. 
The fragmentation that inevitably happens is caused by social balance. 
Societies with weak Heider's balance all become fragmented below a universal 
``critical'' temperature, regardless of their communication densities. 
The phase diagram in Fig. \ref{fig:g001} demonstrates this point. Here we used a very weak social balance of $g = 0.01$. 
The fragmented phase (yellow) extends  as the relative effect of social balance, $g$, increases, as seen in Fig. \ref{fig:phasediag2}. 
Rescaling $g$ to a non-zero value simply shifts the critical line without changing the structure of the phase diagram.

\subsection{Relative update frequencies---choice  of parameter $n$}

\begin{figure}[tb]
\begin{center}
\includegraphics[scale=0.24]{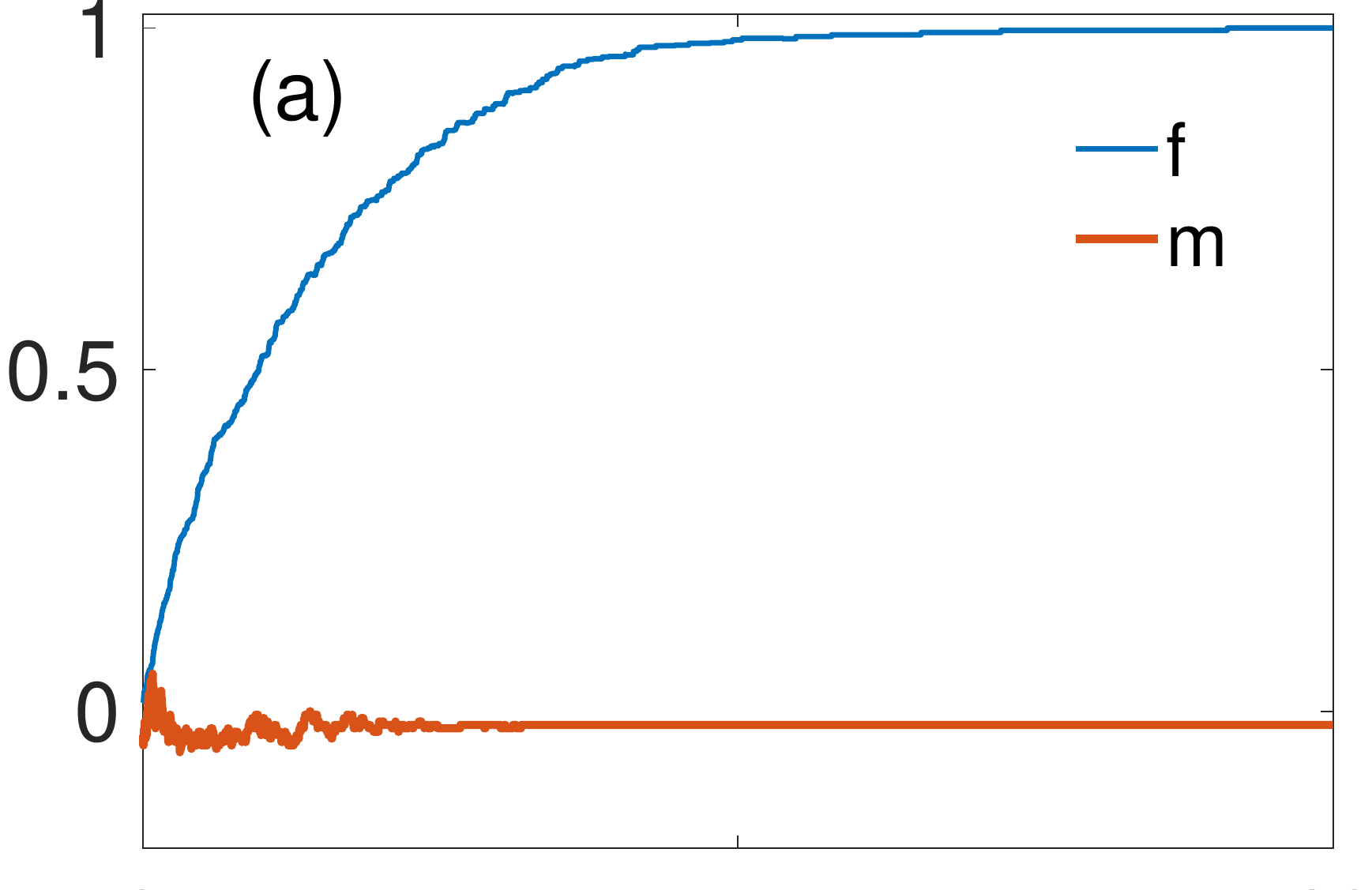}
\, 
\includegraphics[scale=0.24]{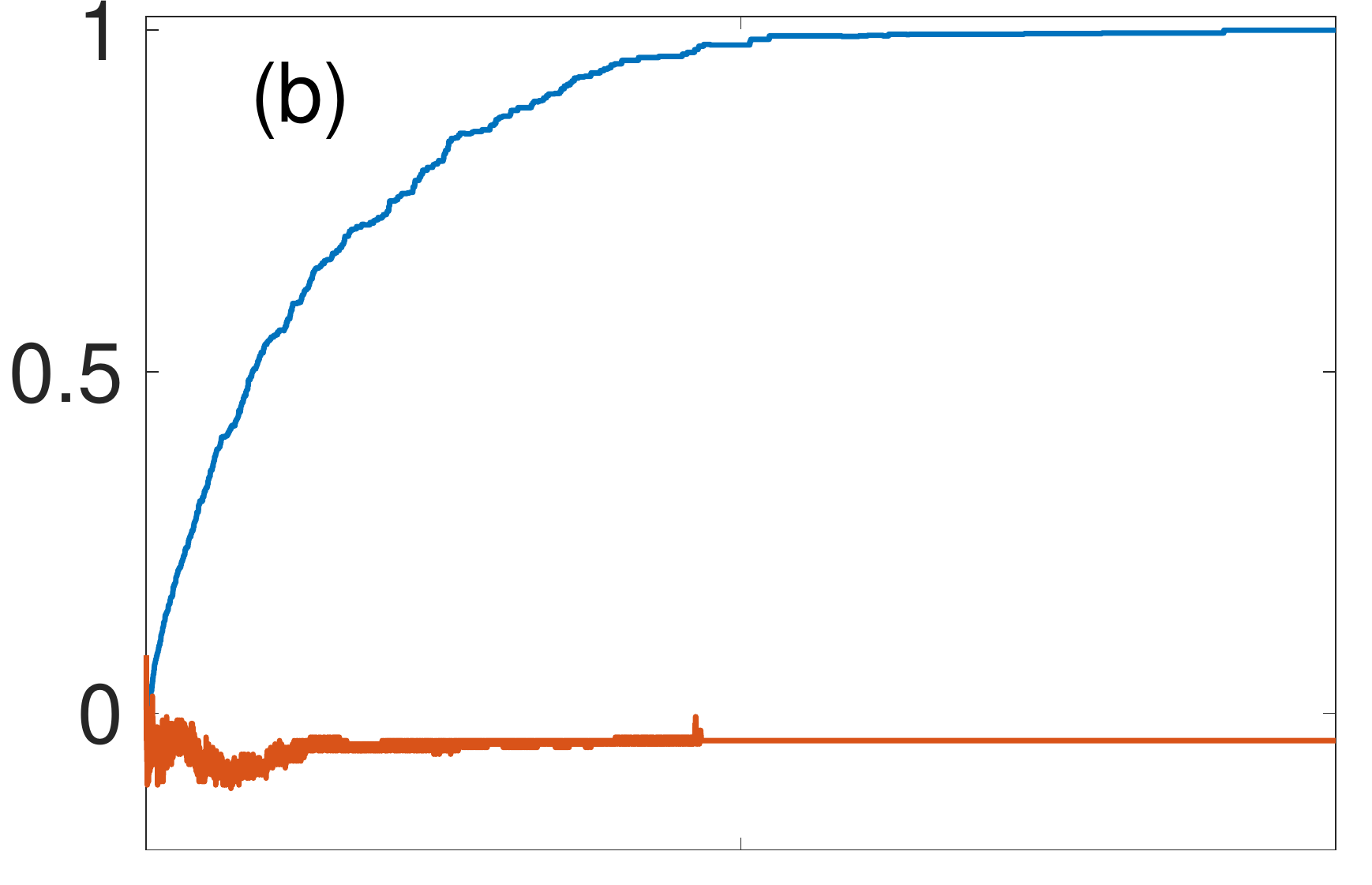}
\includegraphics[scale=0.24]{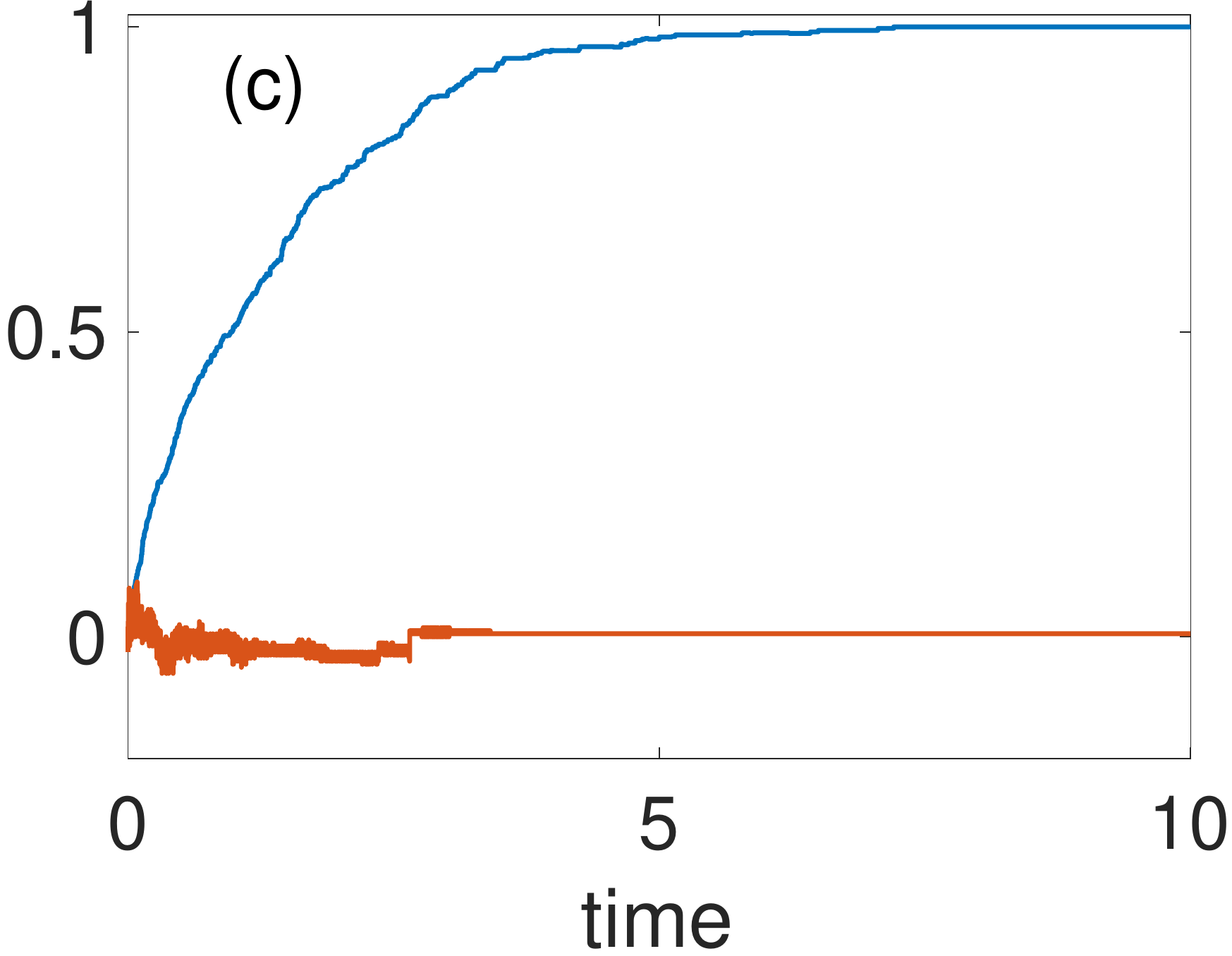}
\includegraphics[scale=0.24]{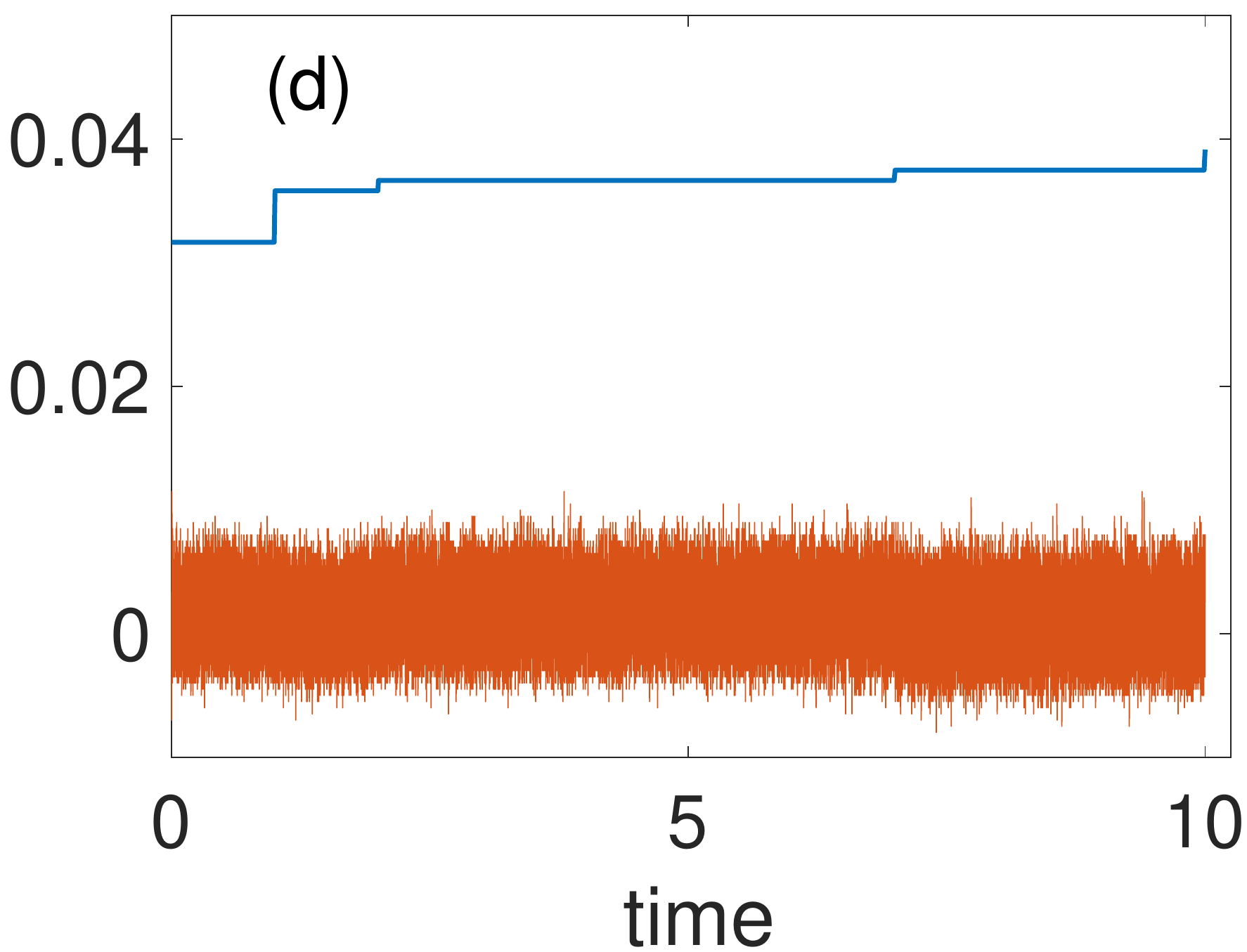}
\end{center}
\caption{Evolution of $f$ and $m$ over time with different  values of $n$: (a) $n = 0.01$,  (b) $n = 1$, (c) $n =100$, and (d) $n = 10,000$. 
$N = 400$, $k = 8$, $T = 1$, $g = 1$, $h = 0$. 
Note that in (d), $m$ has been rescaled by a factor of 10 to keep it within the range of the $y$ axis. 
Time is measured in units of $k N$ timesteps.
}
    \label{fig:evolution2}
\end{figure}

Figure \ref{fig:evolution2}  shows how the observables $f$ (blue) and $m$ (red) evolve over time for different values of $n$. 
For values of $n$ that ensure a correct implementation of the co-evolutionary  dynamics ($n = 0.01, 1, 100$), 
the system evolves in more or less the same way, see panels (a)-(c). 
Based on this observation we made our choice for $n =1$ as used in the paper. 
Only when links evolve very slowly compared to opinions ($n = 10,000$)  do
we observe significant deviations; the system can no longer equilibrate during the simulation time. Note the change of scale in 
panel (d).

\subsection{Illustration of external influences, $h$}
\begin{figure}[h]
\begin{center}
\includegraphics[scale=0.20]{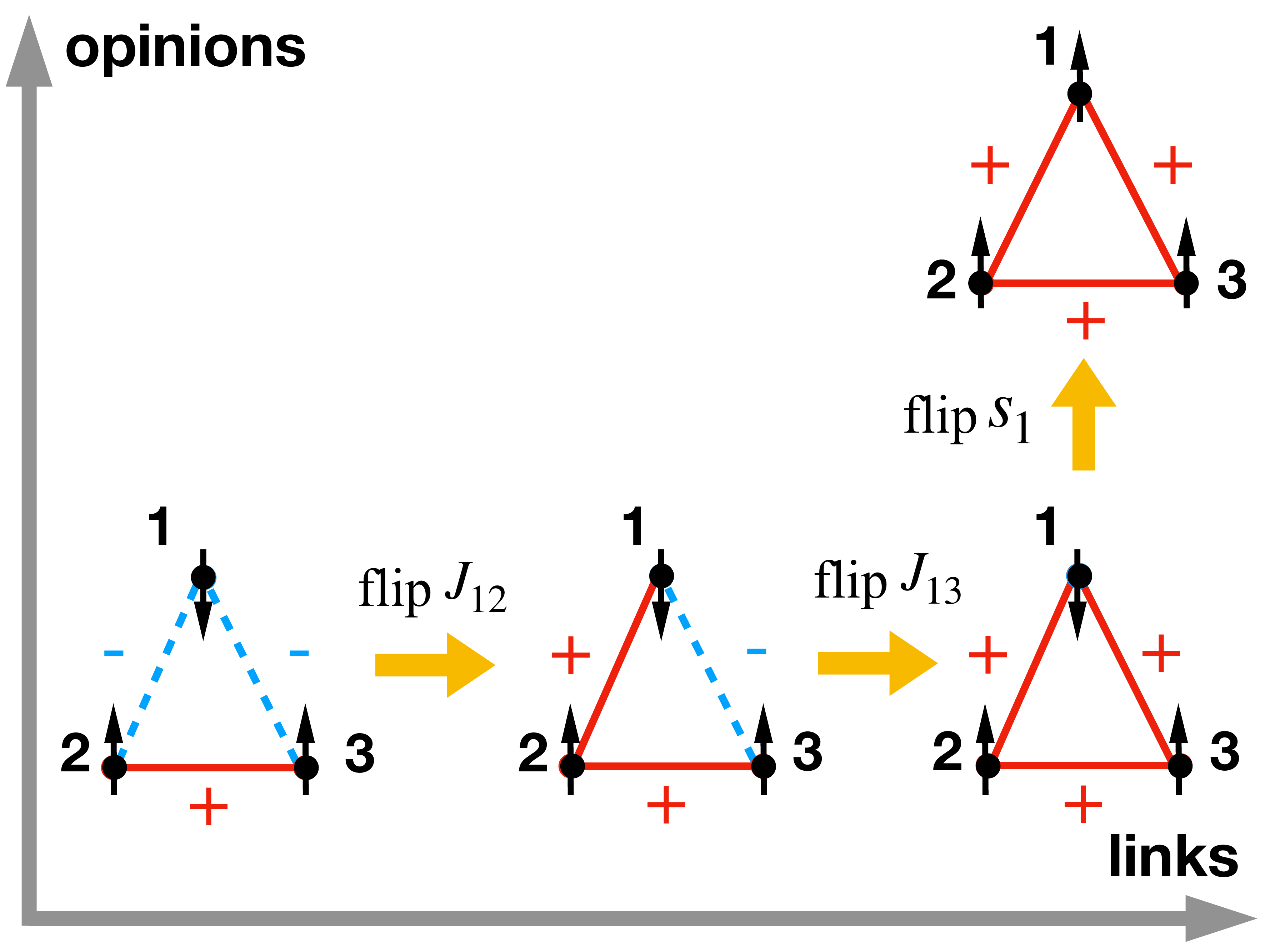}
\end{center} 
\caption{Evolution from a bipolar (bottom-left) to the consensus state (top-right) in the phase space of opinions ($y$-axis) and links ($x$-axis) for $h \geq h_c \equiv 2(1 + g)$ at  zero temperature $T = 0$. 
If $h < h_c$ then a system  in  the basin of attraction of the bipolar state would never end up in the consensus phase at $T = 0$, 
following this path. In this state, the effects of $g$ and $h$ are opposing each other. 
$g$ tends not to flip either of the two negative links, $J_{12}$, since the triangle is already  balanced, 
but $h$ favours its removal to reduce the number of negative links. 
Only a value of $h$ larger than the threshold, $h_c$, can overcome the effect of $g$ and bring the whole system to consensus.
}
    \label{fig:path}
\end{figure}

To illustrate the effect of $h$ and its competing role with $g$ in some cases, 
in Fig. \ref{fig:path} we show  an evolutionary path, which a system of three individuals 
would follow if $h \geq h_c$. For this particular system $h_c = 2(1+g)$ at zero temperature. 
When $h < h_c$, both the  bipolar (bottom-left) and global consensus (top-right) states are  attractors. 
If the system starts from any configuration in the basin of attraction of the bipolar  state, it  ends up in this state. 
This state is no longer the global minimum of the social stress, $H$, if an external influence, $h > h_c$, 
is introduced. The system will then be driven towards consensus, following, for example, 
the illustrated path.

\subsection{Number of triangles in small-world networks}
\begin{figure}[h]
\begin{center}
\includegraphics[scale=0.38]{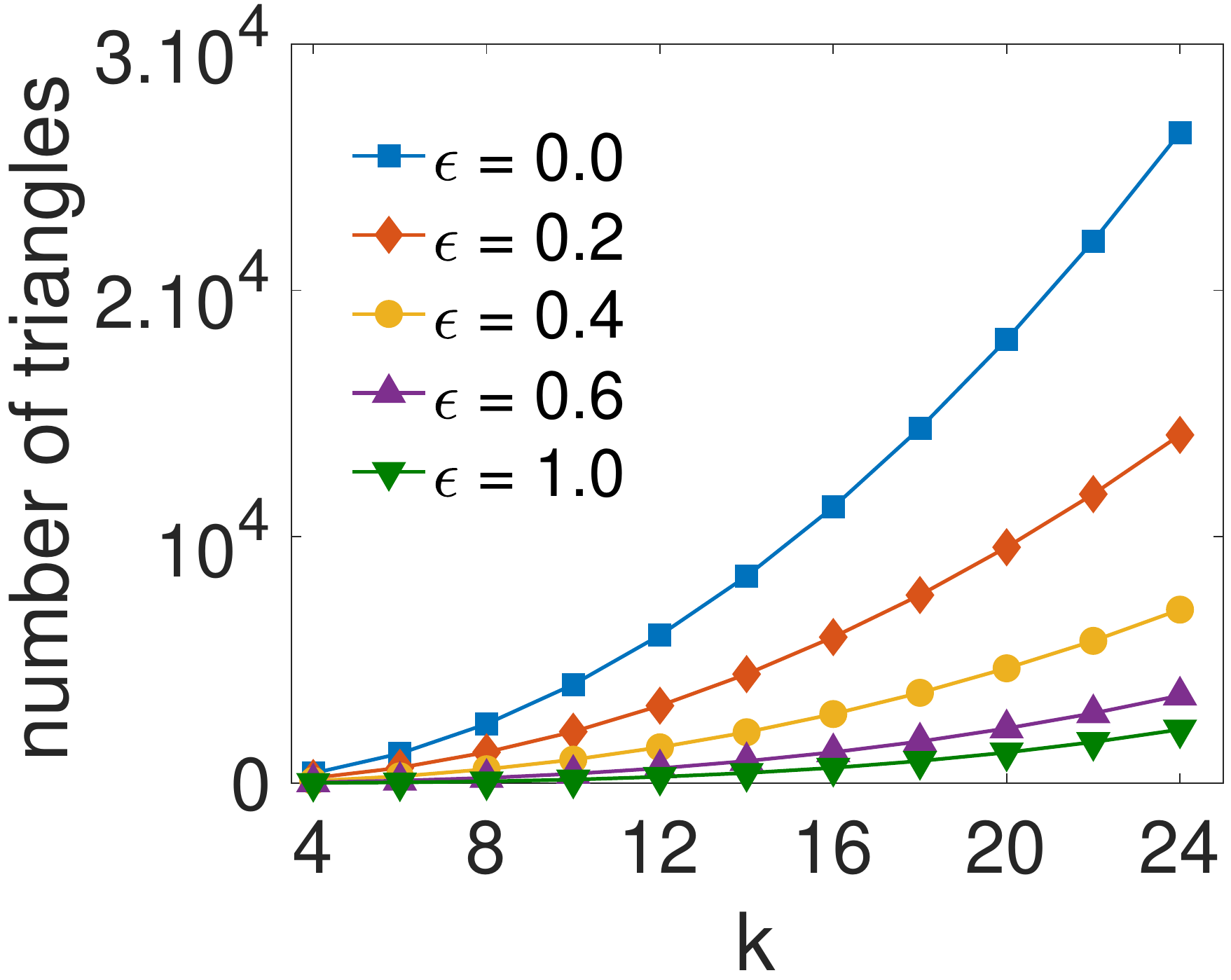}
\end{center}
\caption{Number of  triangles as a function of the network connectivity $k$ for different values of 
the small-world parameter, $\epsilon$. 
$N = 400$, results are averaged over 500 realisations for each value of $\epsilon$.
}
    \label{fig:trianglesinWS}
\end{figure}

Figure \ref{fig:trianglesinWS} shows the number of triangles in small-world networks as a 
function of $k$ and the rewiring parameter, $\epsilon$. 
It is obvious that less triangles exist in networks with more random connections (larger $\epsilon$). 
Therefore, the effect of Heider's balance on the dynamics via triadic relations must become 
weaker as $\epsilon$ increases. This heuristically explains why the critical line is shifted towards the left, 
i.e., the fragmented phase  shrinks with increasing $\epsilon$.

\end{document}